\documentclass{iopart}
\usepackage{latexsym}
\usepackage{ntheorem}
\usepackage{iopams}
\usepackage{setstack}
\usepackage[colorlinks=true]{hyperref}

\newcommand{\What}{{\mathring W}}
\newcommand{\timW}{{\hat  W}}

\newcommand{\genK}{L}%
\newcommand{\genL}{\underbar{{\mbox{\emph L}}}\,}%

\newcommand{\Ker}{{\mathrm{Ker}}}

\newcommand{\og}{{\overline{g}}}

\newcommand{\eeal}[1]{\label{#1}\end{eqnarray}}

\newcommand{\bed}{\begin{deqarr}}
\newcommand{\eed}{\end{deqarr}}
\newcommand{\bedl}[1]{\begin{deqarr}\label{#1}}
\newcommand{\eedl}[2]{\arrlabel{#1}\label{#2}\end{deqarr}}

\newcommand{\tg}{{\widetilde{g}}}

\newcommand{\mcU}{{\mycal U}}

\newcommand{\mcN}{{\mycal N}}

\newcommand{\bel}[1]{\begin{equation}\label{#1}}
\newcommand{\bea}{\begin{eqnarray}}
\newcommand{\bean}{\begin{eqnarray}\nonumber}
\newcommand{\beal}[1]{\begin{eqnarray}\label{#1}}
\newcommand{\eea}{\end{eqnarray}}

\newcommand{\qed}{\hfill $\Box$ \medskip}
\newcommand{\proof}{\noindent {\sc Proof:\ }}
\newcommand{\be}{\begin{equation}}
\newcommand{\eeq}{\end{equation}}
\newcommand{\ee}{\end{equation}}
\newcommand{\beqa}{\begin{eqnarray}}
\newcommand{\eeqa}{\end{eqnarray}}
\newcommand{\beqan}{\begin{eqnarray*}}
\newcommand{\eeqan}{\end{eqnarray*}}
\newcommand{\ba}{\begin{array}}
\newcommand{\ea}{\end{array}}

\newcommand{\mcM}{{\mycal M}}

\newcommand{\mcL}{{\mycal L}}

\newcommand{\beqar}{\begin{deqarr}}
\newcommand{\eeqar}{\end{deqarr}}
\newcommand{\beaa}{\begin{eqnarray*}}
\newcommand{\eeaa}{\end{eqnarray*}}

\DeclareFontFamily{OT1}{rsfs}{}
\DeclareFontShape{OT1}{rsfs}{m}{n}{ <-7> rsfs5 <7-10> rsfs7 <10-> rsfs10}{}
\DeclareMathAlphabet{\mycal}{OT1}{rsfs}{m}{n}

\newtheorem{theorem}{Theorem}[section]

\def\polhk#1{\setbox0=\hbox{#1}{\ooalign{\hidewidth
  \lower1.5ex\hbox{`}\hidewidth\crcr\unhbox0}}}

\providecommand{\MR}{\relax\ifhmode\unskip\space\fi MR }

\providecommand{\href}[2]{#2}

\begin{document}

\title{The many ways of the characteristic Cauchy problem}
\author{Piotr T Chru\'sciel and Tim-Torben Paetz}
\address{Gravitational Physics, University of Vienna, Boltzmanngasse 5, 1090 Vienna, Austria}
\eads{\mailto{Piotr.Chrusciel@univie.ac.at}, \mailto{Tim-Torben.Paetz@univie.ac.at}}

\begin{abstract}
We review various aspects of the characteristic initial value problem for the Einstein equations, presenting new approaches to some of the issues arising.
\end{abstract}
\vspace{-1.5em}


\section{Introduction}

By now there exist four
well established ways of systematically constructing general solutions of the Einstein equations,  by solving
\begin{enumerate}
 \item a spacelike Cauchy problem (see \cite{ChoquetYork79,BartnikIsenberg} and references therein), or
 \item a boundary-initial value problem~\cite{FriedrichNagy,KRSWCMP,FriedrichAdS} (for further references see~\cite{SarbachTiglio}), or
 \item a characteristic Cauchy problem on two transverse hypersurfaces, or
 \item a characteristic Cauchy problem on the light-cone.
\end{enumerate}

One can further consider mixtures of the above.
The aim of this paper is to present some new approaches to the last two questions, and to review the existing ones.

To put things in perspective, recall that Einstein's equations by themselves do not have any type that lends itself directly into a known mathematical framework which would provide existence and/or uniqueness of solutions~\cite{FriedrichRendall}. The monumental discovery of Yvonne Choquet-Bruhat in 1952~\cite{ChBActa} was,  that the imposition of wave-equations on the coordinate functions led to a system where both existence and uniqueness could be proved. The  constraint equations satisfied by the initial data
on a spacelike hypersurface turned out to be both necessary and sufficient conditions for solving the problem. The constraint equations, and the ``harmonicity conditions" became then the two standard notions in our understanding of the spacelike Cauchy problem.

In the early
1960's there arose a strong interest in the characteristic initial value problem
because of  attempts to formulate non-approximate notions of gravitational radiation
in the non-linear theory~\cite{Newman:Penrose,BBM,Penrose,SachsCIVP,Dautcourt}.
While those papers provided much insight into the problem at hand, it is widely recognized that the first mathematically satisfactory treatment of the Cauchy problem on two intersecting null hypersurfaces is due to Rendall~\cite{RendallCIVP}, see also~\cite{CagnacEinsteinCRAS1,CagnacEinsteinCRAS2,ChristodoulouMzH,PenroseCIVP,F2,DamourSchmidt,MzHSeifertCIVP,CaciottaNicoloI,CaciottaNicoloII,F1,HaywardNullSurfaceEquations}. Rendall's initial data consist of a conformal class of a family of two-dimensional
Riemannian
metrics
\begin{equation*}
 \tilde \gamma:= \gamma_{AB}(r,x^C)\mathrm{d}x^A\mathrm{d}x^B
\end{equation*}
on the null hypersurfaces, complemented by suitable data on the intersection. Here $r$ is an affine parameter on the null geodesics threading the initial data surfaces. Rendall uses the Raychaudhuri equation to compute the conformal factor $\Omega$
needed to determine the family of physically relevant data
\bel{28XII11.1}
 \tilde g:= g_{AB}(r,x^C)\mathrm{d}x^A \mathrm{d}x^B \equiv \Omega^2 (r,x^C)   \gamma_{AB}(r,x^C)\mathrm{d}x^A\mathrm{d}x^B
\ee
on the null hypersurfaces. The harmonicity conditions and the Einstein equations determine then uniquely the whole metric $g$ to the future of the initial data surfaces and near the intersection surface $S$. The reader will find more details in Section~\ref{S28XII11.1}.

Rendall's elegant approach works well in vacuum, and more generally  for a class of matter fields that  includes  scalar,  Maxwell, or Yang-Mills fields.
However, it appears  awkward to use an unphysical family of conformal metrics as initial data, since the physically relevant, and geometrically natural, object is the family $\tilde g$. In this context it appears appropriate to view the tensor field  $\tilde g$ as an initial datum on the characteristic surfaces, with the Raychaudhuri equation playing the role of a constraint equation. The idea of prescribing $\tilde \gamma$ should then be viewed as a conformal ansatz for constructing solutions of this constraint equation.

The last point is only a question of interpretation. More importantly, Rendall's scheme does not work for e.g.\ the Einstein-Vlasov equations for particles with prescribed rest mass $m$
(see~\cite{YvonneLiouville} for an existence theorem for those equations with initial data on a spacelike hypersurface),
because the energy-momentum tensor for the Vlasov field,%
\footnote{We have included a factor $8\pi$ in the definition of $T_{\mu\nu}$, so that the Einstein equations read $S_{\mu\nu}=T_{\mu\nu}$, where $S_{\mu\nu}$ is the Einstein tensor.}
\begin{equation}
 \label{19XII11.10}
 T_{\alpha\beta} = 8 \pi \int_{\{g_{\rho\sigma}p^\rho p^\sigma = -m^2\} } f p_\alpha p_\beta\, \mathrm{d}\mu (p)
 \;,
\end{equation}
where $f=f(x,p)$ is the Vlasov distribution function and $\mathrm{d}\mu\equiv\mathrm{d}\mu(p)$ is the Riemannian measure induced on the ``mass-shell" $\{g_{\rho\sigma}p^\rho p^\sigma = -m^2\}$,
depends explicitly upon all components of the metric. This leads to the need of reformulating the problem so that the whole metric tensor is allowed as part of the initial data on the characteristic surfaces. Such a method will be presented in Section~\ref{S28XII11.2}, after having reviewed Rendall's approach in Section~\ref{S28XII11.1}.
We will do this both for data given   on two transversally intersecting null hypersurfaces, and on a light-cone.

We complement the above with a geometric formulation of the characteristic initial data in Section~\ref{s12III12.1},
where we give geometric interpretations of $n$, out of $n+1$, wave-map gauge constraint equations.

The bottom line of our analysis is, that the gravitational characteristic initial data have to satisfy one single constraint equation, the Raychaudhuri  equation. This raises the question, how to construct solutions thereof.
In Section~\ref{solving_constraints} we present several methods to do this.
In the short Sections~\ref{s4V12.20}-\ref{ss22I12.1} we recall how this has already been done in the preceding sections.
In Section~\ref{ss23II12.1} we analyse the Hayward gauge condition $\kappa=\tau/(n-1)$, which may be used as alternative to an affine-parameterization-gauge where the function $\kappa$ vanishes.

It has been proposed to use the shear tensor $\sigma$ as the free initial data for the gravitational field rather than $\tilde \gamma$. However, one defect is that it is not clear how to guarantee tracelessness of $\sigma$.
We  present in Section \ref{sssCfd} a tetrad formulation of the problem to get rid of this grievance.
Finally, we adapt in Section~\ref{ss26I12.1xyz} to any dimensions an approach of Helmut Friedrich (originally developed in dimension four, using spinors), where certain components of the Weyl tensor are used as unconstrained initial data for the gravitational field.
Again, this requires to work in  a null-frame formalism  to take care of the tracelessness of the Weyl tensor.

In the case of a light-cone the Hayward gauge leads us to the issue:
Under which conditions is the assumption, that the vertex is located at the origin $r=0$ of the adapted coordinate system, consistent  with regularity at the vertex?
This question is considered in an appendix.

\section{Rendall's approach}
 \label{S28XII11.1}

In this section we review Rendall's approach to the characteristic initial value problem. For definiteness in the remainder of this section we will consider the vacuum Einstein equations; we comment  at the end of this section on those energy-momentum tensors which are compatible with the analysis here.

Consider two smooth hypersurfaces $N_a$, $a=1,2$, in an $(n+1)$-dimensional
manifold $\mcM $,
with  transverse intersection along a smooth submanifold $S$.
Near the $N_a$'s one can choose adapted coordinates $(x^1,x^2,x^A)$ so that    $N_1$  coincides with the set $\{x^1=0\}$, while $N_2$ is given  by $\{x^2=0\}$.
The hypersurfaces $N_a$ are supposed to be characteristic, which is equivalent to the requirement that, in the coordinates above, on $N_1$ the metric takes the form
\begin{equation}
 \label{29X0.1a0}
 {g}|_{N_1}=\overline{g}_{11}(\mathrm{d}x^{1})^{2}+2\overline{g}_{12}\mathrm{d}x^{1}\mathrm{d}x^{2}
 +2\overline{g}_{1A}\mathrm{d}x^{1}\mathrm{d}x^{A}+
\underbrace{\overline{g}_{AB}\mathrm{d}x^{A}\mathrm{d}x^{B}}_{=:\tilde g}
 \;,
\end{equation}
similarly on $N_2$. Here, and elsewhere, the terminology and notation of \cite{CCM2} is used, in particular an overbar over a quantity denotes restriction to the initial data surface $N_1\cup N_2$.  Rendall  assumes moreover  that $x^2$ is an affine parameter along the curves  $\{x^1=0, x^A=\mathrm{const}^A\}$, and that $x^1$ is an affine parameter along the curves $\{x^2=0, x^A=\mathrm{const}^A\}$.

On $N_1$ let
\begin{equation}
\tau \equiv \frac{1}{2}\overline g^{AB}\partial_2 \overline g_{AB}
\label{dfn_tau}
\end{equation}
be the divergence scalar,
and let
\begin{equation}
 \sigma_{AB} \equiv \frac{1}{2} \partial_2 \overline g_{AB} - \frac{1}{n-1}\tau \overline g_{AB}
\label{dfn_sigma}
\end{equation}
 be the trace-free part of $\partial_ 2  \overline g_{AB}$, also known as the shear tensor.
 The vacuum Raychaudhuri equation,
\begin{equation}
\partial_2\tau  + |\sigma|^2 + \frac{\tau^2}{n-1} = 0
 \;,
  \label{21XII11.30x}
\end{equation}
provides a constraint equation on the family of two-dimensional metrics $x^2\mapsto \overline{g}_{AB}(x^2,x^C)\mathrm{d}x^A\mathrm{d}x^ B$,
where
\begin{equation*}
 |\sigma|^2 \equiv \sigma_A{}^B \sigma_B{}^A \;,
 \quad \sigma_A{}^B \equiv \overline g^{BC} \sigma_{AC}
 \;.
\end{equation*}
Note that $\sigma_A{}^B$ depends only on the conformal class of $\overline g_{AB}$.
 As shown by Rendall, metrics satisfying the constraint (\ref{21XII11.30x}) can be constructed by  freely prescribing the family $x^2\mapsto \gamma_{AB}\mathrm{d}x^A \mathrm{d}x^ B$. Writing $\overline{g}_{AB}= \Omega^2 \gamma_{AB}$, (\ref{21XII11.30x}) becomes then a second order ODE in $x^2$ for $\Omega$,
\begin{eqnarray}
\fl 0  =   \partial^2_{2}\Omega + \frac{\Omega}{n-1}\Big[  \frac{1}{2}\partial_2(\gamma^{AB}\partial_2\gamma_{AB}) +  \underbrace{|\sigma|^2+ \frac{1}{4(n-1)}(\gamma^{AB}\partial_2\gamma_{AB})^2}_{=-\frac{1}{4}\partial_2\gamma^{AB}\partial_2\gamma_{AB}}\Big]
\nonumber
\\
 \label{ODE_Omega}
 + \frac{1}{n-1}\gamma^{AB}\partial_2\gamma_{AB}  \partial_2\Omega
\;.
\end{eqnarray}
This needs to be complemented by $\Omega|_S$ and $\partial_2\Omega|_S$.

Let us require  all coordinate functions to satisfy the scalar wave equation $\Box_g x^{\mu}=0$.
Then the affine parameterization condition $\Gamma^2_{22}|_{N_1}=0$  can be rewritten as
\begin{equation*}
 \partial_2 \overline g_{12} = \frac{1}{2} \tau\overline g_{12}
 \;.
\end{equation*}
This equation determines the metric function $g_{12}$ on $N_1$ with the freedom to prescribe $g_{12}$ on $S$.

The equation   $\overline R_{2A} =0$ on $N_1$ takes the form
\begin{eqnarray}
 - \frac{1}{2}(\partial_2 + \tau)\xi_A  + \tilde{\nabla}_B \sigma_A^{\phantom{A}B} - \frac{n-2}{n-1}\partial_A\tau =0
 \;,
 \label{20XII11.1x}
\end{eqnarray}
where $\tilde{\nabla}$ is the covariant derivative operator of the metric $g_{AB}\mathrm{d}x^A \mathrm{d}x^B$. Here, using the assumption that all coordinate functions satisfy the wave equation, the covector $\xi_A$ reads
(\cite[Equation~(8.25)]{CCM2})
\begin{eqnarray}
 \xi_A &:= & -2\overline g^{12}\partial_2\overline g_{1A} + 4\overline g^{12}\overline g_{1B}\sigma_A{}^B + 2\overline g^{12}\overline g_{1A}\tau  - \overline g_{AB} \overline g^{CD} \tilde \Gamma{}^B_{CD}
 \;,
 \label{20XII11.2x}
\end{eqnarray}
where the $ \tilde \Gamma{}^B_{CD}$'s are the Christoffel symbols of  the metric $\overline g_{AB}\mathrm{d}x^A \mathrm{d}x^B$.
This provides an ODE for the metric functions $\overline{g}_{1A}$: Indeed,
one determines $\xi_A$ by integrating  (\ref{20XII11.1}), with the freedom
to prescribe
\begin{equation*}
 \xi_A^{N_1}:= \xi_A(x^2 =0)
\end{equation*}
on $S$.
(One should keep in mind that   $\xi_A$ here is unrelated to the corresponding  field $\xi_A$ on $N_2$, determined by an analogous equation where all quantities $\tau$, $\sigma$, etc., are calculated using the fields on $N_2$.)

Then $\overline g_{1A}$ is found by integrating (\ref{20XII11.2x}). Since the metric needs to be continuous,  the metric component $g_{1A}$ has to vanish at $S$;
this requirement defines the integration constant.
We further observe that by definition of $\xi_A$ the freedom to prescribe $ \xi_A^{N_1}$  corresponds to the freedom of prescribing $\partial_2 g_{1A}$ on $S$.

The equation $\overline g^{AB} \overline R_{AB} =0$
 in vacuum takes the form
\begin{equation}
(\partial_{2} + \tau) \zeta +\tilde{R}-\frac{1}{2}\bar{g}%
^{AB}\xi_{A}\xi_{B} +\bar{g}^{AB}\tilde{\nabla}_{A}\xi_{B}=0
 \;,  \label{9.2xx}
\end{equation}
where $\tilde R$ is the curvature scalar of $\tilde g$, and where
\begin{equation}
\zeta :=(2\partial_{2}  + \tau )\bar{g}^{22}
 \;.  \label{9.3}
\end{equation}
Taken together, those equations provide
a second order ODE for $\overline{g}^{22}$;  integration requires the knowledge of $\overline{g}^{22}$ and $\partial_2 \overline{g}^{22}$ on $S$.
Employing the relation $\overline g^{22} = (\overline g^{12})^2(\overline g^{AB}\overline g_{1A}\overline g_{1B} - \overline g_{11})$ we observe
that $ g^{22}$ has to vanish at $S$, while there remains the freedom of prescribing $\partial_2 g^{22}$, equivalently
$\partial_2  g_{11}$, on $S$.

However, the validity of the harmonicity conditions
implies certain constraints on $S$ (see below): The value of $\partial_2  g_{11}$ at $S$ is
determined by equation (\ref{SD3y}); similarly the function $\partial_2  g_{11}$ at $S$ follows from (\ref{SD1y}).

One has thus determined all the metric functions $g_{\mu\nu}$ on $N_1$; the procedure on $N_2$ is completely analogous. These are the data needed for the harmonically-reduced Einstein equations, which form a well-posed evolutionary system for the metric.

However, not every solution of the equations constructed in this way will satisfy the vacuum Einstein equations: One still needs to make sure that the harmonicity conditions are satisfied. There is in fact one more subtlety, as one needs to verify that the parameter $x^2$ is indeed an affine parameter on the null geodesics threading $N_1$.
It turns out~\cite{RendallCIVP} that all this will be verified provided three more conditions are imposed at $S$: If we write   $ \nu_A^+$
for what was $ {g}_{1A}|_{N_1}$ so far,
$ \nu_A^-$ for  $ {g}_{2A}|_{N_2}$,   the
wave-coordinates conditions  will hold if we require
that on $S$
\numparts
\begin{eqnarray}
 \partial_1 g_{22}|_S &=& \og_{12} \frac{\partial_2\sqrt{\det g_{AB}}}{\sqrt{\det g_{AB}}}
 \;,
 \label{SD1y}
\\
  \partial_1 \nu_A^- +       \partial_2  \nu_A^+ &= &   \frac{1}{\sqrt{\det g_{EF}}} g_{AB} \partial_C(\og_{12} \sqrt{\det g_{EF}}
        g^{BC})
        \;,
  \label{SD2y}
\\
 \partial_2 g_{11}|_S &=& \og_{12} \frac{\partial_1\sqrt{\det g_{AB}}}{\sqrt{\det g_{AB}}}
  \;.
  \label{SD3y}
\end{eqnarray}
\endnumparts

As already indicated above, the integration functions $\partial_1 g_{22}|_S$ and $\partial_2 g_{11}|_S$ cannot be freely specified but have to be chosen in such a way that
the equations  (\ref{SD1y}) and (\ref{SD3y}) are fulfilled.
Equation (\ref{SD2y}) will be satisfied by exploiting the freedom in the choice of  $\partial_1 \nu_A^-$ and $ \partial_2  \nu_A^+$.
So there remains the freedom to prescribe, say,  $\partial_1 \nu_A^- - \partial_2  \nu_A^+$.

The constraint equation (\ref{SD2y}) can be tied to a terminology introduced by Christodoulou~\cite{ChristodoulouBHFormation} as follows:
Let $\genK$ and $\genL$ be   two null normals to a
codimension-two spacelike hypersurface $S$
satisfying
\begin{equation*}
 g (\genK,\genL)=-2
 \;.
\end{equation*}
Christodoulou~\cite{ChristodoulouBHFormation}
defines the
\emph{torsion one-form} of $S$ by the formula
\bel{tordef}
 \zeta(X) = \frac 12 g(\nabla_X \genK, \genL)
 \;.
\ee
where $X\in TS$. Assuming $g_{12}|_S$ is positive  we can choose $L=\sqrt{2\overline g^{12}}\,\partial_2$ then, on $S$, using the notation above,
$ \genL = -\sqrt{2 \og^{12}}\, \partial_1$,
and (\ref{tordef}) reads
\begin{eqnarray}
 \zeta_A  & = &\frac{1}{2} g(\nabla_A L, \underline L)
 = \frac{1}{2}\overline g^{12}\partial_A \overline g_{12} - \overline \Gamma{}^2_{2A} = \frac{1}{2}(\overline \Gamma{}^1_{1A} - \overline\Gamma{}^2_{2A})
\nonumber
 \\
 &= &\frac{1}{2} \overline g^{12}(\partial_1  \overline  g_{2A} - \partial_2  \overline  g_{1A} ) = \frac{1}{2} \overline g^{12}(\partial_1\nu_A^- - \partial_2\nu_A^+)
 \;.
\label{dfn_zeta_A}
\end{eqnarray}
So $\zeta_A $ contains precisely the information needed to
determine $\partial_1 \nu_A^-$ and $\partial_2 \nu_B^+$ at
$S$, after taking into account (\ref{SD2y}).

\begin{theorem}[Rendall]
  Consider two smooth hypersurfaces $N_1$ and $N_2$ in an $(n+1)$-dimensional manifold with transverse intersection along a smooth submanifold $S$ in
 adapted null coordinates. Let $\gamma_{AB}$ be a smooth family of Riemannian metrics on $N_1\cup N_2$, continuous at $S$.
Moreover, let $\Omega$, $\partial_1\Omega$, $\partial_2\Omega$, $f$
 and $f_A$, $A=3,\dots, n+1$, be smooth fields on $S$,
where we assume $\Omega$ and $f$ to be nowhere vanishing on $S$.
Then there exists an open neighbourhood $U$ of $S$ in the region $\{x^1\geq 0 , x^2 \geq 0\}$, a unique function $\Omega$ on $(N_1\cup N_2)\cap U$ and
a unique smooth Lorentz metric $g_{\mu\nu}$ on $U$ such that
\begin{enumerate}
 \item $g_{\mu\nu}$ satisfies the vacuum Einstein equations,
 \item $\overline g_{AB}=\Omega^2 \gamma_{AB}$,
 \item $\Omega$ induces the given data on $S$, $g_{12}|_S = f$ and $\zeta_A = f_A$.
\end{enumerate}
\end{theorem}

This analysis of the constraints applies equally well to a light-cone with some minor modifications~\cite{CCM2}, where the wave-equations for the coordinate functions are replaced by wave-map conditions. Moreover, there is no need to provide further initial data at the tip of the light-cone, as those are replaced by conditions arising from the requirement of regularity of the metric there; the reader is referred to~\cite{CCM2} for a detailed discussion.
An explicit parameterization of tensors $\tilde g$ which arise by restriction of a smooth metric in normal coordinates has been given in~\cite{ChJezierskiCIVP}. The reader should keep in mind the serious difficulties with regularity of the metric at the vertex, when attempting to prove an existence theorem for the light-cone problem; see~\cite{CCM3,CCM4} for results under restrictive conditions on the data.

The above extends easily to non-vacuum models with  energy-momentum tensors of the form
\begin{eqnarray*}
\fl \overline T_{22} &=& \overline T_{22}(\mathrm{matter \;  data},\gamma_{AB}, \partial_i\gamma_{AB}, \Omega, \partial_2 \Omega, \overline g_{12}, \partial_i \overline g_{12}, x^i)\;, \quad i=2,A\;,
\\
\fl  \overline T_{2A} &=& \overline T_{2A}(\mathrm{matter \; data}, \gamma_{AB}, \partial_i\gamma_{AB}, \Omega, \partial_i \Omega,  \overline g_{1i}, \partial_i \overline  g_{12}, \partial_2 \overline  g_{1A}, \overline{\partial_1 g_{22}}, x^i)\;,
\\
\fl  \overline T_{12} &=& \overline T_{12}(\mathrm{matter \; data},\gamma_{AB}, \partial_i\gamma_{AB}, \Omega, \partial_i \Omega,  \overline g_{1\mu}, \partial_2 \overline  g_{1\mu}, \partial_A \overline g_{12}, \overline{\partial_1 g_{2i}}, x^i)\;,
\end{eqnarray*}
on the initial surface $ \{x^1=0\}$, cf.\ \cite{CCM2}.

\section{All components of the metric as initial data}
 \label{S28XII11.2}

Let $\ell^\nu$ denote the field of null tangents to a characteristic hypersurface.
In this section we present a treatment of the characteristic Cauchy problem which applies to energy-momentum tensors of the form
\begin{equation}
 \label{19XII11.11}
   T_{\mu\nu} \ell^\nu = T_{\mu }  (g,\phi, \partial^\parallel g, \partial ^\parallel \phi, x)
   \;,
\end{equation}
for some fields $\phi$ satisfying equations which, when the metric is considered as given, possess a well-posed characteristic Cauchy problem. Here the symbol $\partial^\parallel $ denotes derivatives in directions purely tangential to the initial data surfaces.
In particular (\ref{19XII11.11}) includes the Einstein-Vlasov case.

As already discussed, in~\cite{RendallCIVP} the corresponding problem for the vacuum Einstein equations is solved using an affine parameterisation of the generators and a wave-map (``harmonic") gauge.
In Rendall's approach some components of the metric are calculated by solving the characteristic-harmonic gauge constraint equations, which
form a hierarchical ODE-system  along the generators of the initial surface.
For an energy-momentum tensor (\ref{19XII11.10})  this approach will generally lead instead to a quasi-linear PDE-system for the metric components. To establish an existence result for that system might be intricate, if possible at all.
It is in any case not obvious how to include an energy-momentum tensor (\ref{19XII11.10}) in this scheme, compare~\cite{CalvinEV}.
We  circumvent the problem by using a gauge adapted to the initial data, where the metric tensor, and thereby  (\ref{19XII11.10}), is fully given on the initial surface, while
the \emph{wave-gauge source vector}  ${\mathring W}{}^{\mu}$
is computed from the values of the metric on the initial surface using the \emph{Einstein -- wave-map-gauge} constraint equations of~\cite{CCM2}.

We start with an analysis of two intersecting hypersurfaces,   the case of a  light-cone will be covered in Section~\ref{ss10I12.1}.

\subsection{Two transverse hypersurfaces}
 \label{ss19XII11.1}

Consider two smooth hypersurfaces $N_a$, $a=1,2$, in an $(n+1)$--dimensional manifold $\mcM $, with  transverse intersection along a smooth submanifold $S$.
As before, we choose adapted null coordinates $(x^1,x^2,x^A)$ so that    $N_1$  coincides with the set $\{x^1=0\}$, while $N_2$ is given  by $\{x^2=0\}$. We use a ``generalized wave-map gauge" as in \cite{CCM2}, with
target metric $\hat g$ of the form
\begin{equation*}
 \hat g = 2\mathrm{d}x^1\mathrm{d}x^2 +  \hat g_{AB}(x^1,x^2,x^C)\,\mathrm{d}x^A\mathrm{d}x^B
 \;.
\end{equation*}
Here $\hat g_{AB}$ is any family of Riemannian metrics on $S$ parameterized by $x^1$ and $x^2$, smooth in all variables. The metric $\hat g$ is only introduced so that the harmonicity vector $H^\mu$, defined in equation (\ref{WaveGauge0x}),  is a vector field, and plays no significant role in what follows.

As gravitational initial data  on the initial  hypersurfaces we prescribe all metric components $\overline{g}_{\mu\nu}$ in the coordinates above, as well as a connection coefficient $\kappa$; this needs to be supplemented by initial data $\overline \phi$ for $\phi$. For instance, in the Einstein-Vlasov case, the supplementary data will be a function $\overline{f}$ defined on the mass-shell $\{\overline{g}_{\mu\nu}p^\mu p^\nu =-m^2\}$, viewed as a subset of the pull-back of $T\mcM $ to the $N_a$'s.

The hypersurfaces $N_a$ are supposed to be characteristic, which is equivalent to the requirement that, in the coordinates above, on $N_1$ the metric takes the form
\begin{equation}
 \label{29X0.1a}
 {g}|_{N_1}=\overline{g}_{11}(\mathrm{d}x^{1})^{2}+2\overline{g}_{12}\mathrm{d}x^{1}\mathrm{d}x^{2}
 +2\overline{g}_{1A}\mathrm{d}x^{1}\mathrm{d}x^{A}+
\underbrace{\overline{g}_{AB}\mathrm{d}x^{A}\mathrm{d}x^{B}}_{=:\tilde g}
 \;,
\end{equation}
similarly on $N_2$. Here, and elsewhere, the terminology and notation of \cite{CCM2} is used, in particular an overbar over a quantity denotes restriction to the initial data surface $N_1\cup N_2$. (Some obvious renamings need to be applied to the equations in \cite{CCM2}, for instance the variable $u$ there is $x^1$ on $N_1$, and $x^2$ on $N_2$; the variable $r$ there is $x^2$ on $N_1$ and $x^1$ on $N_2$.)

We want to apply Rendall's existence theorem~\cite{RendallCIVP} for an appropriately reduced system of equations. For this the trace, $\overline g_{\mu\nu}$, of the metric on the initial surface $N_1\cup N_2$ needs to be
the restriction of a smooth Lorentzian spacetime metric.
This will be the case if $\overline g_{12}$ is nowhere vanishing, if  $\overline g_{AB}|_{N_a}$ is a family of Riemannian metrics, and if $\overline g_{\mu\nu}$  is
smooth on $N_1$ and $N_2$ and continuous across $S\equiv N_1\cap N_2$.
We therefore need to impose the following continuity conditions on $S$,
\numparts
\begin{eqnarray}
  \lim_{x^2\rightarrow 0} g_{AB}|_{N_1}&=\, \lim_{x^1\rightarrow 0}g_{AB}|_{N_2} \;,&
 \label{regularity1a}
\\
  \lim_{x^2\rightarrow 0}g_{12}|{N_1} &=\,   \lim_{x^1\rightarrow 0}g_{12}|{N_2}  \;,&
 \label{regularity1b}
\\
  \lim_{x^2\rightarrow 0}g_{1A}|{N_1}&=\,   0  \;, \quad   \lim_{x^1\rightarrow 0} g_{2A}|{N_2} &=\, 0
 \;,
 \label{regularity1c}
\\
 \lim_{x^2\rightarrow 0} g_{11}|{N_1} &=\, 0  \;, \quad  \lim_{x^1\rightarrow 0}g_{22}|{N_2}&=\, 0
 \;.
 \label{regularity1d}
\end{eqnarray}
\endnumparts

Let $H^\mu$ be the \emph{harmonicity vector}, defined as
\begin{equation}
H^{\lambda} :=
g^{\alpha\beta} \Gamma_{\alpha\beta}^{\lambda}-W^{\lambda}
\;, \enspace \mathrm{with}\enspace
W^{\lambda}:=
 \underbrace{g^{\alpha\beta} \hat{\Gamma}_{\alpha \beta}^{\lambda}}_{=:\timW^\lambda} + \What^{\lambda}
\;,
\label{WaveGauge0x}
\end{equation}
where $\What^\lambda$ will be a vector depending only upon the coordinates, and determined by the initial data in a way to be described below.
(In principle $\What^\lambda$ can be allowed to depend on the metric as well, but not on derivatives of the metric.)
To obtain a well posed system of evolution equations we will impose the generalized wave-map gauge condition
\begin{equation*}
 H^{\lambda}=0
   \;.
\end{equation*}
More precisely, we view the wave-map gauge constraints~\cite{CCM2}  as equations for the restriction  $\overline{\What }{}^\mu$ of $\What{}^\mu$ to $N_1$ and $N_2$.
We will solve those equations
hierarchically. We emphasize that, assuming (\ref{19XII11.11}), all components of the energy momentum tensor restricted to $N_1$ and $N_2$ are explicitly known since $\overline g_{\mu\nu}$ and $\overline \phi$ are.

We present the calculations on $N_1$, the equations on $N_2$ are obtained by interchanging the index 1 with the index 2 in all the formulae.

Let $S_{\mu\nu}$ denote the Einstein tensor.
In the notation and terminology of~\cite{CCM2} the first constraint, arising from the equation $\overline S_{22}\equiv \overline R_{22}  =\overline  T_{22}$ evaluated on $N_1$, reads (see \cite[Equation~6.11]{CCM2})
\begin{equation}
   -\partial_2\tau +  { \kappa}\tau - |\sigma|^2 - \frac{\tau^2}{n-1} = \overline  T_{22}
 \;,
  \label{21XII11.30}
\end{equation}
where $\tau$ and $\sigma$ are defined as in (\ref{dfn_tau}) and (\ref{dfn_sigma}), respectively.
Indeed, using the formulae in \cite[Appendix~A]{CCM2} one finds
\begin{eqnarray}
l  \overline{\partial_1\Gamma^1_{22}} &= &\partial_2\overline\Gamma{}^1_{12} + (\overline\Gamma{}^1_{12})^2-  \overline\Gamma{}^1_{12}\overline\Gamma{}^2_{22}
 \qquad\quad\Longrightarrow\quad
 \nonumber
\\
   \overline S_{22} &= & \overline{\partial_1\Gamma^1_{22}} - \partial_2(\overline\Gamma{}^1_{12} -\overline\Gamma{}^A_{2A}) + (\overline\Gamma{}^1_{12}+\overline\Gamma{}^A_{2A})\overline\Gamma{}^2_{22} - (\overline\Gamma{}^1_{12})^2 -\overline\Gamma{}^A_{2B}\overline\Gamma{}^B_{2A}
 \nonumber
\\
  &= &- \partial_2\overline\Gamma{}^A_{2A} + \overline\Gamma{}^A_{2A}\overline\Gamma{}^2_{22} -\overline\Gamma{}^A_{2B}\overline\Gamma{}^B_{2A}
 \nonumber
\\
  &=& -\partial_2 \tau + \tau \overline\Gamma{}^2_{22} - \chi_A{}^B\chi_B{}^A
 \;,
 \label{S22}
\end{eqnarray}
where
\begin{equation*}
 \chi_A{}^B := \frac{1}{2}\overline g^{BC}\partial_2\overline g_{AC}
 \;.
\end{equation*}

\emph{Here we adapt the point of view that the function $\kappa$ is the value on $N_1$ of the Christoffel coefficient $\Gamma^2_{22}$, and is part of the initial data.} Hence, we view (\ref{21XII11.30}) as a constraint equation linking $\overline g_{AB}$, $\kappa$, and the matter sources (if any).

In the region where $\tau$ has no zeros, (\ref{21XII11.30}) can be trivially solved for $\kappa$ to give
\begin{eqnarray}
 \kappa &=&\frac{\partial_2\tau + \frac{1}{n-1}\tau^2 + |\sigma|^2 + \overline T_{22}}{\tau}
 \;.
  \label{20XII11.21a}
\end{eqnarray}
It follows that $\kappa$ does not need to be included as part of initial data when $\tau$ has no zeros, and then the constraint equation (\ref{21XII11.30}) can be replaced by the last equation, determining $\kappa$.

Equation (\ref{20XII11.21a}) can still be used, by continuity, to determine $\kappa$ on the closure of the set where $\tau$ has no zeros, keeping in mind that the requirement of smoothness of the function so determined imposes non-trivial constraints on the right-hand-side. In any case (\ref{20XII11.21a}) does not make sense if there are open regions where $\tau$ vanishes. It seems therefore best to assume that $\kappa$ is any smooth function on $N_1$ such that (\ref{21XII11.30}) holds, and view that last equation as a constraint equation relating $\kappa$, $\overline g _{AB}$ and its derivatives, and the matter fields; similarly on $N_2$.

It should be kept in mind that, once a candidate  solution of the Einstein equations has been constructed,
one needs to verify that $\kappa$ is indeed the value of $\Gamma^2 _{22}$ on $N_1$. We will return to this in (\ref{23II12.1}).

We choose $\overline \What{}^ 1$ to be
\begin{eqnarray}
 \overline \What {}^1 &:=& -\overline{\timW }{}^1 -\overline g^{12}(2\kappa+\tau) - 2 \partial_2\overline g^{12}
  \;.
  \label{20XII11.21}
\end{eqnarray}
Note that the right-hand side is known, so this defines $ \overline \What {}^1$. By definition, this is the $\partial_1$ component of $\What^\mu$ in the coordinate system $(x^1,x^2,x^A)$. We will define the remaining components of $\What^\mu$ shortly, the resulting collection of functions transforming \emph{by definition} as a vector when changing coordinates.

From \cite[Appendix~A]{CCM2} one then finds
\bel{23II12.10}
 \overline \Gamma{}^2 _ {22} = \kappa - \frac 12 \overline{g_{12}}\overline H{} ^1
 \;,
\ee
and
 (\ref{S22}) together with (\ref{21XII11.30}) give
\bel{23II12.2}
 \overline{S}_{22} -\overline{T}_{22} = - \frac 12  \overline{g_{12}}\overline H{} ^1 \tau
 \;.
\ee

The corresponding constraint equation on $N_2$ determines $\What {}^2|_{N_2}$. We shall return to  the question of continuity at $S$ of $\What{}^1|_{N_1\cup N_2}$ and of $\What{}^2 |_{N_1\cup N_2}$ shortly.

The next constraint equation follows from $\overline S_{2A}  \equiv \overline R_{2A} =\overline  T_{2A}$.
From the formulae in \cite[Appendix~A]{CCM2} we find
\begin{eqnarray*}
 \overline{\partial_1\Gamma^1_{2A}} &=& \partial_A\overline\Gamma{}^1_{12} +\overline\Gamma{}^1_{12}(\overline\Gamma{}^1_{1A}-\overline\Gamma{}^2_{2A}) + \overline\Gamma{}^B_{12}\overline\Gamma{}^1_{AB} -\overline\Gamma{}^B_{2A}\overline\Gamma{}^1_{1B}
 \;,
\end{eqnarray*}
which gives
\begin{eqnarray}
\fl  \overline S_{2A} =\overline{\partial_1\Gamma^1_{2A}} + \partial_2\overline\Gamma{}^2_{2A} + \partial_B\overline\Gamma{}^B_{2A} - \partial_A\overline\Gamma{}^1_{12} - \partial_A\overline\Gamma{}^2_{22} - \partial_A\overline\Gamma{}^B_{2B} + \overline\Gamma{}^1_{1B}\overline\Gamma{}^B_{2A}
 \nonumber
\\
  + \overline\Gamma{}^1_{12}(\overline\Gamma{}^2_{2A} - \overline\Gamma{}^1_{1A}) + \overline\Gamma{}^B_{2B}\overline\Gamma{}^2_{2A} + \overline\Gamma{}^B_{BC}\overline\Gamma{}^C_{2A} - \overline\Gamma{}^1_{AB}\overline\Gamma{}^B_{12} - \overline\Gamma{}^B_{AC}\overline\Gamma{}^C_{1B}
 \nonumber
\\
 \fl  \phantom{ \overline S_{2A}}=\partial_2 \overline\Gamma{}^2_{2A} + \partial_B\overline\Gamma{}^B_{2A}  - \partial_A\overline\Gamma{}^2_{22} - \partial_A\overline\Gamma{}^B_{2B} + \overline\Gamma{}^B_{2B}\overline\Gamma{}^2_{2A} + \overline\Gamma{}^B_{BC}\overline\Gamma{}^C_{2A} -  \overline\Gamma{}^B_{AC}\overline\Gamma{}^C_{2B}
 \nonumber
\\
 \fl  \phantom{ \overline S_{2A}} = (\partial_2 + \tau)\overline\Gamma{}^2_{2A} + \tilde\nabla_B\chi_A{}^B - \partial_A\overline\Gamma{}^2_{22} - \partial_A\tau
 \;,
 \label{S2A}
\end{eqnarray}
where $\tilde\nabla$ is the covariant derivative associated to the Riemannian metric $g_{AB}$.
That leads us to the equation
\begin{eqnarray}
   - \frac{1}{2}(\partial_2 + \tau)\xi_A  + \tilde{\nabla}_B \sigma_A^{\phantom{A}B} - \frac{n-2}{n-1}\partial_A\tau -\partial_A \kappa
 = \overline T_{2A}
 \;,
 \label{20XII11.1}
\end{eqnarray}
where the field $\xi_A$ denotes the restriction of the (rescaled) Christoffel coefficient $-2 \Gamma{}^2_{2A}$ to $N_1$.
We determine  $\xi_A$ by integrating  (\ref{20XII11.1}), with the freedom
to prescribe
\begin{equation*}
 \xi_A^{N_1}:= \xi_A(x^2 =0)
\end{equation*}
on $S$.  (One should keep in mind that   $\xi_A$ here is unrelated to the corresponding  field $\xi_A$ on $N_2$, determined by an analogous equation where all quantities $\tau$, $\sigma$, etc., are calculated using the fields on $N_2$.)
We then define $\overline{\What}{}^A$ through the formula
\begin{eqnarray}
 \overline \What {}^A &:=&\overline g^{AB}\Big[\xi_B + 2\overline g^{12} (\partial_2\overline g_{1B} - 2 \overline g_{1C}\sigma_B{}^C -\overline g_{1B} \tau)  -\overline g_{1B}(\overline \What {}^1 + \overline{\timW }{}^1 ) \Big]
  \nonumber
\\
&&   +\overline g^{CD}\tilde\Gamma^A_{CD} - \overline{\hat W}{}^A
 \;;
  \label{23II12.5}
\end{eqnarray}
equivalently
\begin{eqnarray}
   \xi_A =  -2\overline g^{12}\partial_2\overline g_{1A} + 4\overline g^{12}\overline g_{1B}\sigma_A{}^B + 2\overline g^{12}\overline g_{1A}\tau + \overline g_{1A}(\overline \What{}^1+\overline{\timW }{}^1)
 \nonumber
\\
   + \overline g_{AB}(\overline \What{}^B + \overline{\timW }{}^B) - \overline g_{AB} \overline g^{CD} \tilde \Gamma{}^B_{CD}
 \;.
 \label{20XII11.2}
\end{eqnarray}
This has been chosen so that, using the formulae in \cite[Appendix~A and Section~9]{CCM2},
\begin{equation}
\overline{S}_{2A}-\overline{T}_{2A} =
-\frac{1}{2}(\partial_{2}+\tau)(\og_{AB}\overline{H}{}^{B}+\og_{1A}\overline{H}{}^{1}) +\frac{1}{2}\partial_{A}(\og_{12}\overline{H}{}^{1})
\; .
\label{LAfinal}
\end{equation}
Moreover, one finds (cf.\ equation (10.35) in \cite{CCM2})
\begin{equation}
 \xi_A = -2\overline\Gamma{}^2_{2A} - \overline g_{AB} \overline H{}^B -  \overline{g}_{1A} \overline H{}^1
 \;.
 \label{relation_xiA_HA}
\end{equation}

On $S$ (\ref{23II12.5}) takes the form
\begin{eqnarray*}
 \overline \What {}^A |_S&=&\overline g^{AB}\Big[\xi_B^{N_1} +  2\overline g^{12} \partial_2\overline g_{1B}\Big] +\overline g^{BC}(\tilde\Gamma^A_{BC} - \hat\Gamma^A_{BC})
 \;.
\end{eqnarray*}
Keeping in mind the corresponding equation on $N_2$,
\begin{eqnarray*}
 \overline \What {}^A |_S&=&\overline g^{AB}\Big[\xi_B^{N_2} +  2\overline g^{12} \partial_1\overline g_{2B} \Big] +\overline g^{BC}(\tilde\Gamma^A_{BC} - \hat\Gamma^A_{BC})
 \;.
\end{eqnarray*}
the requirement of continuity of $\What{}^A|_{N_1 \cup N_2}$ leads to
\begin{equation}
 \label{21XII11.5}
 \xi_A^{N_1} - \xi_A^{N_2}
     =
    2  g^{12} (   \partial_1 g_{2A}  -  \partial_2 g_{1A} )|_S \equiv 4 \zeta_A
  \;.
\end{equation}
Recall that the torsion one-form $\zeta_A$ has been defined in (\ref{dfn_zeta_A}).

We continue with the equation $\overline S_{12}   = \overline T_{12}$, or, equivalently,
\[\overline g^{AB}\overline R_{AB} = -(2\overline g^{12} \overline T_{12}+\overline g^{22}\overline T_{22} + 2\overline g^{2A}\overline T_{2A})= \overline g^{AB} \overline T_{AB} -\overline T  \;,
\]
which we handle in a manner similar to the previous equations. Using the identities (10.33) and (a corrected version of
\footnote{On the right-hand-side of (10.36) in \cite{CCM2}, in the conventions and notations there, a term $\tau \overline g^{11}/2$ is missing.}%
)
(10.36)  in \cite{CCM2} we find that on $N_1$ we have
\begin{eqnarray*}
\fl \overline g^{AB} \overline R_{AB}  \equiv (\partial_2 + \overline \Gamma{}^2_{22} + \tau)(2\overline g^{AB} \overline\Gamma{}^2_{AB} +\tau\overline g^{22}) - 2\overline g^{AB} \overline\Gamma{}^2_{2A} \overline\Gamma{}^2_{2B}
 - 2 \overline g^{AB} \tilde\nabla_A \overline\Gamma{}^2_{2B}  + \tilde R
 \;,
\end{eqnarray*}
and we are led to the equation
\begin{eqnarray}
 (\partial_2 + \kappa + \tau) \zeta   + \big( \tilde\nabla_A - \frac{1}{2}\xi_A\big) \xi^A + \tilde R =   \overline g^{AB} \overline T_{AB} -  \overline T
 \;,
 \label{constraint3}
\end{eqnarray}
with $\xi^A:=\overline g^{AB} \xi_B$, and
where the quantity $\zeta$ denotes the restriction of
\begin{equation*}
  2 \big(\overline g^{AB} \overline\Gamma{}^2_{AB}+ \tau \overline g^{22}
 \big)
\end{equation*}
to $N_1$.
We integrate  (\ref{constraint3}),  viewed as a first-order ODE for $\zeta$.
The initial data on $S$   are determined by the requirement of continuity of $\overline\What{}^2  $ at $S$, which we choose to be
\begin{equation}
\overline \What {}^2  :=   \frac 12 \zeta  - (\partial_{2}+\kappa + \frac 12 \tau )\bar{g}^{22}-\overline{\timW }{}^2\;.  \label{9.3x}
\end{equation}
Indeed, recall that $\overline \What {}^2|_S $ has already been calculated algebraically when analysing the  first constraint equation  on $N_2$, in exactly the same way as we calculated $\overline \What {}^1$ in the first step of the analysis above.

Similarly the  initial data for the integration of the constraint which determines $\What{}^1|_{N_2}$ is determined by the requirement of continuity of $\What{}^1|_{N_1\cup N_2}$.

The choice  (\ref{9.3x}) has been done so that
\begin{eqnarray}
\fl  \overline g^{AB} \overline R_{AB} - \overline g^{AB} \overline T_{AB} +\overline T \,=\, (\partial_2 +\kappa +  \tau -\frac{1}{2}\overline g_{12} \overline H{}^1 ) (2\overline H{}^2  - \overline g_{12}\overline g^{22}\overline H{}^1 )    -\frac{1}{2}\zeta\overline g_{12} \overline H{}^1
 \nonumber
\\
  +(\tilde\nabla_A- \xi_A - \frac{1}{2}\overline g_{AB}\overline H{}^B - \frac{1}{2}\overline g_{1A}\overline H{}^1)( \overline H{}^A + \overline g_{1C} \overline g^{AC}\overline H{}^1)
 \;.
 \label{L0final}
\end{eqnarray}
We note that our choice of $\overline \What {}^2$ is equivalent to
\begin{eqnarray}
 \zeta= 2\overline g^{AB}\overline \Gamma{}^2_{AB} + \tau \overline g^{22}  +  \overline g_{12}\overline g^{22}\overline H{}^1 - 2 \overline H{}^2
 \;.
 \label{relation_zeta_H2}
\end{eqnarray}

Summarising, given the fields $\kappa$, $\overline g_{\mu\nu}$ and   $\overline\phi$ on $N_1\cup N _2$,
 satisfying
 (\ref{21XII11.5}), and the sum $\xi^{N_1}_A+\xi^{N_2}_A$ on $S$, we have found a unique continuous
vector field  $\overline{\What}$ on  $N_1\cup N _2$, smooth up-to-boundary on $N_1$ and $N_2$,
 so that  (\ref{23II12.2}),  (\ref{LAfinal}) and  (\ref{L0final}) hold  on $N_1\cup N _2$. Letting $\What$ be any smooth vector field on $\mcM $ which coincides with $\overline\What$ on $N_1\cup N _2$, and assuming that  the reduced Einstein equations (see  (\ref{RicciH}) below) can be complemented by well-posed evolution equations for the matter fields, we obtain a metric, solution~of the Cauchy problem for the reduced Einstein equations in a future neighbourhood of~$S$.

However, the metric so obtained will solve the full Einstein equations if and only if~\cite{CCM2}  $H^\mu$ vanishes on $N_1\cup N_2$, so we need  to ensure that this condition holds. Note that at this stage a smooth metric $g$, satisfying the reduced Einstein equations, and a smooth vector field $W^{\mu}$
are known to the future of $N_1\cup N_2$   in some neighbourhood of $S$, and thus $H^\mu$ is a known smooth vector field there.

By \cite[Section 7.6]{CCM2}, $\overline H {}^1$ will vanish on $N_1$ if and only if $\overline H {}^1$ vanishes on $S$.
Using (\ref{20XII11.21}) and  (\ref{23II12.10}) together with the equations in \cite[Appendix~A]{CCM2} we find
\begin{equation}
 \label{21XII11.1}
  H^1|_S = (g^{12})^2 \partial_1 g_{22} + 2 g^{12} \kappa + 2 \partial_2 g^{12}
  \;.
\end{equation}
We conclude that $  H{}^1|_{N_1}$ will vanish if and only if the initial data $\overline g_{22}$ on $N_2$ have the property that the derivative $\partial_1 g_{22}$ on $S$ satisfies
\begin{equation}
 \label{21XII11.2}
 \partial_1 g_{22}|_S =    2\big(   \partial_2 g_{12} -g_{12} \kappa_{N_1}  \big) \quad \Longleftrightarrow \quad \Gamma^2_{22}|_S = \kappa_{N_1}
  \;,
\end{equation}
where, to avoid ambiguities, we denote by $\kappa _{N_a}$ the function $\kappa$ associated with the hypersurface $N_a$, etc.
Similarly $  H{}^2|_{N_2}$ will vanish if and only if we choose $g_{11}$ on $N_1$ so that
\begin{equation}
 \label{21XII11.3}
 \partial_2 g_{11}|_S =   2  \big( \partial_1 g_{12} -g_{12} \kappa_{N_2}  \big)  \quad \Longleftrightarrow \quad   \Gamma^1_{11}|_S = \kappa_{N_2}
  \;.
\end{equation}
With those choices we have $H^1|_{S}=  H^2|_{S}=0$, and the arguments in~\cite{CCM2} show that $H^1|_{N_2}=  H^2|_{N_1}=0$ as well.

We continue with $\overline H^A$.
Then by equation  (\ref{relation_xiA_HA}) we have
\begin{eqnarray*}
 \xi_A^{N_1} &=& -(2\Gamma^{2}_{2A} + g_{AB} H^B + g_{1A}H^1)|_S
\\
 &=& -\big(\overline{g}^{12}(\partial_A \overline{g}_{12}-\overline{\partial_1g_{2A}}+ \partial_2\overline{g}_{1A})  + g_{AB} H^B + g_{1A}H^1\big)|_S
  \;,
\\
 \xi_A^{N_2} &=&  -(2\Gamma^{1}_{1A} + g_{AB} H^B + g_{1A}H^1)|_S
\\
 &=&-\big(\overline{g}^{12}(\partial_A \overline{g}_{12}-\overline{\partial_2g_{1A}}+ \partial_1\overline{g}_{2A}) + g_{AB} H^B + g_{1A}H^1\big)|_S
  \;,
\end{eqnarray*}
and the conditions $H^A|_S=0$ and $H^1|_S=0$ determine $\xi_A^{N_a}$ in terms of the remaining data. Note that (\ref{21XII11.5}) is then automatically satisfied, and that we loose the freedom to prescribe $ \xi_A^{N_1}+ \xi_A^{N_2}$.

It remains to show that our choice of the parameterization of the null rays is consistent, i.e.\ we have to make sure that the relations $\Gamma^2_{22}|_{N_1}=\kappa_{N_1}$ and $\Gamma^1_{11}|_{N_2}=\kappa_{N_2}$ hold.
This follows trivially from the vanishing of the wave gauge vector $H$ due to the identities
\begin{eqnarray}
 \label{23II12.1}
  H^1|_{N_1} \equiv 2g^{12} (\kappa- \Gamma^2_{22}) \quad \mathrm{and} \quad  H^2|_{N_2} \equiv 2g^{12} (\kappa- \Gamma^1_{11})
 \;.
\end{eqnarray}
Similarly, the vanishing  of $\overline H{}^1$ and  $\overline H{}^A$ shows via (\ref{relation_xiA_HA}) that the identification of $\xi_A$ with the rescaled Christoffel coefficient $-2\overline \Gamma{}^2_{2A}$ on $N_1$ and  $-2\overline \Gamma{}^1_{1A}$ on $N_2$ is consistent.
The vanishing of  $\overline H{}^1$ and  $\overline H{}^2$, together with the identity  (\ref{relation_zeta_H2}) imply that  on $N_1$ the field $\zeta$ indeed represents the value of $2\overline g^{AB} \overline\Gamma{}^2_{AB} +  \tau \overline g^{22}$,
and the corresponding field on $N_2$ represents the value of $2\overline g^{AB} \overline\Gamma{}^1_{AB} +  \tau \overline g^{11}$ there.

In particular, the above provides a new and simple integration scheme for the vacuum Einstein equations, where all the metric functions are freely prescribable on $N_1\cup N _2$:

\begin{theorem}
Given any continuous  functions $(\kappa,\overline g_{\mu\nu})$ on $N_1\cup N_2$ such that
\numparts
\begin{eqnarray}
 \label{29X0.1a1}
 g|_{N_1}=\overline{g}_{22}(\mathrm{d}x^{2})^{2}+2\overline{g}_{12}\mathrm{d}x^{1}\mathrm{d}x^{2}
 +2\overline{g}_{2A}\mathrm{d}x^{2}\mathrm{d}x^{A}+
 \overline{g}_{AB}\mathrm{d}x^{A}\mathrm{d}x^{B}
 \;,
\\
 \label{29X0.1a2}
 g|_{N_2}=\overline{g}_{11}(\mathrm{d}x^{1})^{2}+2\overline{g}_{12}\mathrm{d}x^{1}\mathrm{d}x^{2}
 +2\overline{g}_{1A}\mathrm{d}x^{1}\mathrm{d}x^{A}+
 \overline{g}_{AB}\mathrm{d}x^{A}\mathrm{d}x^{B}
 \;,
\end{eqnarray}
\endnumparts
smooth up-to-boundary on $N_1$ and $N_2$,
and satisfying  (\ref{21XII11.2})-(\ref{21XII11.3}) together with the vacuum constraint equations (here $\kappa_{N_1}:= \kappa|_{N_1}$, etc.)
\numparts
 \begin{eqnarray}
  -\partial_2\tau_{N_1} +  { \kappa _{N_1}} \tau_{N_1} - |\sigma _{N_1}|^2 - \frac{\tau^2 _{N_1}}{n-1} = 0 
   \ \mbox{on $N_1$}
 \;,
  \label{21XII11.40}
 \\
  -\partial_1\tau_{N_2} +  { \kappa _{N_2}} \tau_{N_2} - |\sigma _{N_2}|^2 - \frac{\tau^2 _{N_2}}{n-1} =0 
   \ \mbox{on $N_2$}
 \;,
  \label{21XII11.41}
\end{eqnarray}
\endnumparts
there exists a smooth metric defined on some neighbourhood of $S$, solution of the vacuum
 Einstein equations to the future of $N_1\cup N_2$.
\end{theorem}

Note that all the conditions are necessary. To see this, let $g$ be any metric solving the Einstein equations to the future of $N_1\cup N_2$, with $N_a$ characteristic. We can introduce adapted coordinates near $N_1\cup N_2$ so that (\ref{29X0.1a1})-(\ref{29X0.1a2}) hold. The constraints (\ref{21XII11.40})-(\ref{21XII11.41}) follow then from the Einstein equations~\cite{CCM2}, while (\ref{21XII11.2})-(\ref{21XII11.3}) follow from our calculations above.

\medskip

\proof
While the main elements of the proof have already been given, to avoid ambiguities we summarize the argument:
Let $(\kappa,\overline g)$ be given as above. Set
\begin{equation*}
 \mcM:=[0,\infty)\times [0,\infty)\times S
 \;,
\end{equation*}
where the first $[0,\infty)$ factor refers to the variable $x^1$, and the second to $x^2$. On $\mcM$ let $\hat g$ be the metric $\hat g =2 dx^1 dx^2 + \phi_{AB}dx^A dx^B$, where $\phi_{AB}dx^A dx^B$ is a Riemannian metric on $S$.
Let $\overline{\mathring W}{}^\mu$ be constructed as above.
Let $W^\mu$ be any smooth extension of $\overline{  W}{}^\mu$
 to  $\mcM$, and let $g$ be the solution of the wave-map reduced Einstein equations  $ R_{\alpha\beta}^{(H)}=0$, with initial data $\overline g$, where
\begin{equation}
R_{\alpha\beta}^{(H)}:=R_{\alpha\beta} -{\frac{1}{2}}(g_{\alpha\lambda
}\hat{D}_{\beta}H^{\lambda}+g_{\beta\lambda}\hat{D}_{\alpha}H^{\lambda
}) ,
\label{RicciHIdentity}
\end{equation}
with $H^\mu$ defined by  (\ref{WaveGauge0x}),
and where
$\hat{D}$  is  the Levi-Civita covariant derivative in the metric
$\hat{g}$.
(It follows from~\cite[page 163]{YCB:GRbook} that $R_{\alpha\beta}^{(H)}$ is a
quasi-linear, quasi-diagonal operator on $g$, tensor-valued,
depending on $\hat{g}$, of the form
\begin{equation}
R_{\alpha\beta}^{(H)}\equiv-{\frac{1}{2}}g^{\lambda\mu}
{\hat{D}}_{\lambda} {\hat{D}}_{\mu}g_{\alpha\beta}+
\hat{f}[g,{\hat{D}}g]_{\alpha\beta}\;,
\label{RicciH}
\end{equation}
where $\hat{f}[g,{\hat{D}}g]_{\alpha\beta}$  is a tensor quadratic in $\hat Dg$
with coefficients  depending upon  $g$,  $\hat{g}$,    $\What$,  $\hat D \timW$ and $\hat D \What$; existence of solutions of this problem follows  from~\cite{RendallCIVP}.) As $\overline H^\mu =0$ by construction, a standard argument shows that $H^\mu\equiv 0$, and so $g$ is a solution of the vacuum Einstein equations in a suitable neighbourhood of $S$ in $\mcM$.
\qed

\subsection{The light-cone}
 \label{ss10I12.1}

Now let us consider the same problem on a light-cone $C_O$ with vertex $O$.
We prescribe the metric functions $\overline g_{\mu \nu}$ on the cone in adapted null coordinates (cf.\ \cite{CCM2}) as well as $\overline\phi$ and, if $\tau$ has zeros, $\kappa$
(note that $\tau\equiv \frac{1}{2} \overline g^{AB} \partial_1 \overline g_{AB}$ has no zeros in a sufficiently small neighbourhood around the vertex).
In this section the notations and conventions from \cite{CCM2} are used again; in particular the $x^1$-coordinate will be frequently denoted by $r$, and the light-cone is given as the surface $\{x^0 \equiv u=0\}$.

For $C_O$ to be a characteristic cone we need to have $\overline g_{11}=0=\overline g_{1A}$ in our adapted coordinates.
To end up with a Lorentzian metric, the component $\nu_0\equiv \overline g_{01}$ has to be nowhere vanishing, while $\overline g_{AB}$ has to be a family of Riemannian metrics on $S^{n-1}$.
We consider initial data which satisfy
\numparts
\begin{eqnarray}
 \overline g_{00} \,=\, -1 + O(r^2) \;, \quad  &\partial_r\overline g_{00} \,=\, O(r)
 \;,
 \label{cone_assumption1}
\\
 \nu_0 \,=\, 1 + O(r^2) \;, \quad   &\partial_r\nu_0\,=\, O(r)
 \;,
 \label{cone_assumption2}
\\
 \nu_A \,=\, O(r^3) \;, \quad  &\partial_r \nu_A \,=\, O(r^2)
 \;,
 \label{cone_assumption3}
\\
 \overline g_{AB} \,=\, r^2s_{AB} +  O_2(r^4) \;, \quad &\partial_r\partial_C \overline g_{AB} \,=\, 2r\partial_C s_{AB} + O_1(r^3)
 \;,
 \label{cone_assumption4a}
\\
 \partial_r^2\partial_C\partial_D \overline g_{AB} \,=\, 2\partial_C\partial_Ds_{AB} & \hspace{-0.3em}+ O(r^2)\;,
 \label{cone_assumption4}
\end{eqnarray}
\endnumparts
%
for small $r$, where   $f=O_n(r^\alpha)$ means that $\partial_r^i \partial_A^\beta f = O(r^{\alpha-i}) $ for $i+|\beta| \le n$.
The tensor $s_{AB}$ denotes the round sphere metric.

These conditions ensure that the metric is of the same form near the vertex as in \cite{CCM2}.
The assumptions concerning the derivatives, which are compatible with the relations (4.41)-(4.51) in \cite{CCM2},
 are made to compute the behaviour of $\overline{\mathring W}$ near the vertex:%
\footnote{It is conceivable that a larger class of initial data turns out to be compatible with regularity at the vertex.}
Though we do not attempt to tackle the regularity problem at the vertex here,
as a necessary condition we want to make sure that $\overline{\mathring W}$ remains bounded near the vertex,
 which in our adapted coordinates means
\begin{eqnarray*}
 \overline{\mathring W}{}^0 = O(1)\;, \quad  \overline{\mathring W}{}^1 = O(1)\;, \quad \overline{\mathring W}{}^A = O(r^{-1})
 \;.
\end{eqnarray*}

In fact it turns out that with (\ref{cone_assumption1})-(\ref{cone_assumption4}) and the subsequent assumptions on the target metric and the energy momentum tensor the vector
 $\overline{\mathring W}$  goes to zero.

We present the scheme for an arbitrary target metric $\hat g$ that satisfies the relations
\numparts
\begin{eqnarray}
 \hat \nu_0 &=& 1+  O_1(r^2)\;, \quad \hat\nu_A \,= \,O_1(r^3) \;, \quad \overline{\hat g}_{00} \,= \, -1 + O(r^2)\;,
 \label{target1}
\\
  \partial_r\overline{ \hat g}_{00}  &= & O(r)\;, \quad \overline{\hat g}_{AB}\, =\, r^2 s_{AB} + O_1(r^4)\;,
  \label{target2}
\\
 \overline {\partial_0 \hat g_{11}} &=& O(r)
 \;,
\quad
 \overline {\partial_0 \hat g_{1A}}  \,= \,O(r^2)
 \;,
\quad
 \overline g^{AB} \overline {\partial_0 \hat g_{AB}}  \,= \, O(r)
  \label{target3}
 \;.
\end{eqnarray}
\endnumparts
Again, these assumptions are to ensure that the behaviour of $\overline{\mathring W}$ can be determined at the vertex.

 Additionally, we take a look at two particular target metrics: a Minkowski target $\hat g = \eta$ as in \cite{CCM2}, and a target metric $\hat g = C$ which satisfies $\overline C= \overline g$ and which simplifies the expressions for the components of  $\overline{\mathring W}$.

Let us now solve the constraint equations.
The first constraint yields (supposing that $\tau$ has no zeros, the case where it does have zeros can be treated as in the case of two transversally intersecting hypersurfaces)
\begin{eqnarray}
 \kappa &=&\frac{\partial_1\tau + \frac{1}{n-1}\tau^2 + |\sigma|^2 + \overline {T}_{11}}{\tau}
 \label{relation_kappa-tau}
\end{eqnarray}
and
\begin{eqnarray*}
 \overline \What {}^0 &=& -\overline{\timW }{}^0- \nu^0(2\kappa+\tau) - 2 \partial_1\nu^0
  \;.
\end{eqnarray*}

If we assume
\footnote{These assumptions on the energy-momentum tensor, as well as those which will be made later, will hold for a  tensor $T_{\mu\nu}$ which has bounded components in coordinates which are well behaved near the vertex; note that the $(u,r,x^A)$ coordinates are singular at the vertex.}
\begin{eqnarray*}
\overline T_{11} = O(1)\;, \quad \partial_A\overline T_{11} = O(1)\;,  \quad \partial_A\partial_B\overline T_{11} = O(1)\;,
\end{eqnarray*}
we obtain with our assumptions (\ref{cone_assumption1})-(\ref{cone_assumption4}) and with the assumptions (\ref{target1})-(\ref{target2}) concerning the target metric
\begin{equation*}
 \kappa \,=\, O(r)\;, \quad \partial_A\kappa = O(r) \;, \quad \partial_A\partial_B\kappa = O(r)\;,
\end{equation*}
and
\begin{equation*}
  \overline \What {}^0  \,=\, O(r)
 \,.
\end{equation*}

Let us write $\overset{\eta}{=}$ for an equality which holds when $\hat g $ is the Minkowski metric, with an obvious similar meaning for $\overset{C}{=}$.
Then
\begin{equation*}
\overline{\timW }{}^0 \overset{\eta}{=} -r \overline g^{AB} s_{AB}\;,
\end{equation*}
and also
\begin{equation*}
  \overline \What {}^0 \,\overset{C}{=}\, 2\nu^0(\hat\Gamma^1_{11}- \kappa)
 \;.
\end{equation*}

From the second constraint equation one first determines $\xi_A$.  Recall that this is a first-order ODE. The integration constant which arises is determined by the requirement of finiteness of $\xi_A$
at the vertex (cf.\ \cite[Section~9.2]{CCM2}),
\begin{eqnarray}
 \label{26I12.2}
 \xi_A &\,=\,&
 \displaystyle 2\frac{e^{-\int_1^r(\tau-\frac{n-1}{\tilde r})\mathrm{d}\tilde r}}{r^{n-1}}\int_0^r \tilde r^{n-1} e^{\int_1^{\tilde r}(\tau-\frac{n-1}{\tilde {\tilde r}})\mathrm{d}\tilde{ \tilde r}}
 \nonumber
\\
 && \times \left( \tilde\nabla_B\sigma_A{}^B - \frac{n-2}{n-1}\partial_A\tau - \partial_A \kappa - \overline T_{1A}  \right)\mathrm{d}\tilde r
 \;.
\end{eqnarray}
If we assume
\begin{equation*}
\overline T_{1A} = O(r)\;, \quad \partial_B \overline T_{1A} = O(r)\;,
\end{equation*}
and employ (\ref{cone_assumption1})-(\ref{target3}) we find
\begin{equation*}
 \xi_A \,=\, O_1(r^2)
 \;.
\end{equation*}
The function $\overline \What {}^A$ can then be computed algebraically,
\begin{eqnarray*}
 \fl\overline {\mathring W}{}^A &=& \overline g^{AB} \xi_B + 2\nu^0\overline g^{AB}(\partial_1\nu_B - 2\nu_C\chi_B{}^C)   -\nu_B\overline g^{AB}(\overline {\mathring W}{}^0 + \overline {\hat W}{}^0)
 + \overline g^{BC} \tilde\Gamma^A_{BC} -  \overline{\hat W}{}^A
\\
\fl  &=& O(1)
 \;,
\end{eqnarray*}
where
\begin{equation*}
 \chi_A{}^B \equiv \frac{1}{2}\overline g^{BC} \partial_1 \overline g_{AC}
 \;.
\end{equation*}
In particular
\begin{eqnarray*}
  \overline{\hat W}{}^A&\overset{\eta}{=}& -\frac{2}{r}\nu^0\overline g^{AB}\overline g_{0B} + \overline g^{BC} S^A_{BC}
 \;,
\end{eqnarray*}
and
\begin{eqnarray*}
  \overline{\mathring W}{}^A&\overset{C}{=}& 2\overline g^{1A}(\hat\Gamma^1_{11}-\kappa) + \overline g^{AB}(2 \hat\Gamma^1_{1B} +  \xi_B )
 \;.
\end{eqnarray*}
The functions $S^A_{BC}$ denote the Christoffel coefficients associated to the round sphere metric.

Finally, we have a first-order equation for
\begin{equation}
 \zeta =  (2\partial_1 + 2\kappa + \tau)\overline g^{11} + 2\overline{\mathring W}{}^1 + 2 \overline{\hat W}{}^1
 \;.
 \label{dfn_zeta}
\end{equation}
It reads
\begin{eqnarray}
\fl(\partial_1 + \kappa + \tau)\zeta + \tilde R + \overline g^{AB}(\tilde{\nabla}_A \xi_B - \frac{1}{2}\xi_A\xi_B)
 + \overline g^{11} \overline T_{11} + 2\overline g^{1A}\overline T_{1A}
 + 2\nu^0\overline T_{01} =0
 \;.
\label{30I12.1}
\end{eqnarray}
This can be integrated,
\begin{eqnarray*}
   \zeta &=& \frac{ e^{-\int_1^r (\kappa+\tau-\frac{n-1}{\tilde r})\mathrm{d}\tilde r}}{r^{n-1}}\Big[ c - \int_{0}^r \tilde r^{n-1} e^{\int_1^{\tilde r} (\kappa+\tau-\frac{n-1}{\tilde{\tilde r}})\mathrm{d}\tilde{\tilde r}} \Big( \tilde R \Big. \Big.
\\
 &&\Big.\Big. + \overline g^{AB}\tilde{\nabla}_A \xi_B  - \frac{1}{2}\overline g^{AB}\xi_A\xi_B + \overline g^{11} \overline T_{11} + 2\overline g^{1A}\overline T_{1A} + 2 \nu^0\overline T_{01} \Big) \mathrm{d}\tilde r\Big]
  \;,
\end{eqnarray*}
where $c$ is an integration constant.

Using again the relations (\ref{cone_assumption1})-(\ref{cone_assumption4}) and our assumptions on the target metric we deduce that
\begin{eqnarray*}
\overline g^{11} &=& 1 + O(r^2) \;, \quad \partial_1 \overline g^{11}\,=\, O(r)
 \;,
\\
 \tilde R  &=& (n-1)(n-2) r^{-2} + O(1)
 \;.
\end{eqnarray*}
Assuming that $\overline T_{01} = O(1)$ we find that a general solution $\zeta$ has a term of order~$r^{-(n-1)}$ due to which $\overline{\mathring W}{}^1$ would not converge at the vertex. We thus set $c=0$.
That yields
\begin{eqnarray*}
 \zeta = -(n-1) r^{-1} + O(1)
 \;.
\end{eqnarray*}
Inserting this result into (\ref{dfn_zeta}) we end up with
\begin{equation*}
  \overline \What {}^1 \,=\, O(r)
 \;.
\end{equation*}
For that we employed
\begin{equation*}
  \overline{\timW }{}^1  \,=\, -(n-1) r^{-1} + O(r)
 \;.
\end{equation*}
In the special case of a Minkowski target we have
\begin{equation*}
  \overline{\timW }{}^1 \overset{\eta}{\equiv} \overline{\timW }{}^0  \overset{\eta}{\equiv} -r\overline g^{AB} s_{AB}
\end{equation*}
Moreover, we find
\begin{eqnarray*}
   \overline{\mathring W }{}^1  \overset{C}{=} \frac{1}{2}\zeta  - \overline g^{AB} \hat\Gamma^1_{AB} -\frac{1}{2}\tau \overline g^{11}+ \overline g^{11}(\hat\Gamma^1_{11} - \kappa)
\;.
\end{eqnarray*}

Let us assume that the vector field $ \overline \What {}^{\lambda}$ can be extended to a smooth spacetime vector field $\What {}^{\lambda}$ on the space-time manifold $\mcM$.
If we further assume, as in the case of two transversally intersecting null hypersurfaces, that
the reduced Einstein equations can be complemented by well-posed evolution equations for the matter fields, for sufficiently well behaved initial data we obtain~\cite{DossaAHP} a solution of the Cauchy problem in a future neighbourhood of the tip of the cone.
The metric obtained this way solves the full Einstein equations if and only if  $H$ vanishes on $C_O$, as shown in Sections 7.6, 9.3 and 11.3 of~\cite{CCM2}.

Let us assume now that  initial data $(\kappa,\overline g_{\mu\nu})$ and a target metric $\hat g$ have been specified.
In order to prove that the  wave-map gauge vector  $ H{}^{\lambda}$ vanishes on the cone, one first establishes that it is bounded near the vertex.
In our adapted coordinates that means
\begin{equation}
 \overline H{}^0=O(1)\;, \quad \overline H{}^1 = O(1) \;, \quad \overline H{}^A=O(r^{-1})\;.
 \label{28II12.3}
\end{equation}
If we assume that those transverse derivatives which appear in the generalized wave-map gauge condition $\overline H=0$ satisfy (compare~\cite{CCM2} for a justification under the conditions there)%
\footnote{Note that these transverse derivatives are obtained from the solution $g$ of the reduced Einstein equations with initial data $\overline g$.
The assumptions  (\ref{28II12.3}) are known to hold e.g.\ if one uses the wave-map gauge $\mathring W^{\mu}=0$ near the vertex \cite{CCM2}.
}
\begin{equation*}
 \overline {\partial_0 g_{11}}= O(r)\;, \quad \overline{\partial_0 g_{1A}} = O(r^2)\;, \quad \overline g^{AB} \overline{\partial_0 g_{AB}} = O(r)\;,
\end{equation*}
and the initial data fulfill, additional to  (\ref{cone_assumption1})-(\ref{cone_assumption4}), the relations,
\begin{eqnarray*}
 \partial_A \nu_0 = O(r^2)\;, \quad \partial_B \nu_A = O(r^3)
 \;,
\end{eqnarray*}
then one finds (using  (\ref{cone_assumption1})- (\ref{target3})), say in vacuum,
\begin{equation*}
 \overline H^0=O(r)\;, \quad \overline H^1 = O(r) \;, \quad \overline H^A=O(1)\;,
\end{equation*}
which more than suffices for  (\ref{28II12.3}).

\section{A geometric perspective}
 \label{s12III12.1}

\subsection{One constraint equation}
 \label{s26I11.1}

Let us present a more geometric description of initial data on a characteristic surface.

A triple $(\mcN,\tilde g,\kappa)$ will be called a \emph{characteristic initial data set} if  $\mcN $ is a smooth $n$-dimensional manifold, $n\ge 3$, equipped with a degenerate quadratic
form $\tilde g$ of signature $(0,+,\ldots,+)$, as well as a
connection form $\kappa$ on the one-dimensional degeneracy bundle $\mathrm{Ker }\, \tilde g$, understood as a bundle above its own integral curves.
 The data are moreover required to satisfy a constraint equation, as follows:

We can always locally introduce an adapted coordinate system where  $\mathrm{Ker }\,
\tilde g$ is $\mathrm{Span }\,
\partial_1$.%
\footnote{This vector was  denoted as $\partial_r$ or $\partial_1$ in Section~\ref{ss10I12.1}, by $\partial_1$ in Section~\ref{S28XII11.1} when considering the null hypersurface $ N_2$, and as $\partial_2$ in Section~\ref{S28XII11.1} when considering the null hypersurface $ N_1$.}
   (There only remains the freedom of coordinate transformations of the form $(x^1,x^A) \mapsto (\bar x^1 (x^1, x^A), \bar x^B (x^A))$.)
Then the connection form $\kappa$ reduces to one connection coefficient:
\bel{20I12.1}
 \nabla_{\partial_1} \partial_1 = \kappa \partial_1
 \;.
\ee
In this coordinate system we have $\tilde g =
\og_{AB}\mathrm{d}x^A\mathrm{d}x^B$.
Denoting by
${\og }^{AB}$ the matrix inverse to $ \og _{AB} $,
set
\begin{equation}
\chi_B{}^A:=
 \frac{1}{2} {\og }^{AC}\partial_{1} {\og }_{CB}
 \;,
 \quad
 \tau :=
  \chi_A{}^A
 \;.
\label{chi2x}
\end{equation}
Under redefinitions of the adapted coordinates the field $\tau$ transforms as a covector, which leads to the natural
covariant derivative operator
\bel{29XII11.1}
\nabla_1 \tau:=  (\partial_{1}
-\kappa )\tau
 \;.
\ee
With those definitions,    the \emph{characteristic constraint equation} reads
\begin{equation}
 \label{31III10.1}
 \nabla_1 \tau= - \chi_B{}^A\chi_A{}^B -
 \rho
  \enspace
  \Longleftrightarrow
  \enspace
 \nabla_1 \tau+ \frac 1 {n-1}\tau^2= - \sigma_B{}^A\sigma_A{}^B -
 \rho
 \;,
\end{equation}
where $\rho$ represents the component $  T_{11}|_{\mcN }$ of
the energy-momentum tensor of the associated space-time $({\mcM },g)$ and, as before, $
\sigma$ is the trace-free part of $\chi$.

A triple  $(\mcN, \tilde g, \kappa)$ satisfying (\ref{chi2x}) with
$\rho=0$ will be called \emph{vacuum characteristic initial data}.

An \emph{initial data set on a light-cone} is a characteristic data set where $\mcN$ is a star-shaped neighbourhood of the origin in $\mathbb{R}^n$ from which the origin has been removed, with the tangents to the half-rays from the origin lying in the kernel of $\tilde g$,
 and with $(\tilde g,\kappa)$ having specific behaviour at the origin as described e.g.\ in~\cite{CCM2}.

The reader is referred to~\cite{galloway-nullsplitting} for a clear discussion of the geometry of null hypersurfaces, and to~\cite{JKCRMP,Jezierski:2003hh,JKCPRD} for a further analysis of the objects involved.

\subsection{Dim-$\mcN$ constraint equations}
 \label{ss26I12.1}

An alternative geometric point of view, closely related to that in~\cite{JKCPRD}, is a slight variation of the above, as follows: Instead of considering a connection on the degeneracy bundle $\Ker \,\tg$, viewed as a bundle over the integral curves of $\Ker \,\tg$, one considers a connection on this bundle \emph{viewed as a bundle over $\mcN$}. For this one needs the connection coefficients $\kappa$ and $\xi_A$,
defined by the equations:
\bel{26I12.1}
 \nabla_{\partial_1} \partial_1 = \kappa \partial_1\;, \quad
 \nabla_{\partial_A} \partial_1 = -\frac 12\xi_A \partial_1 + \chi_A{}^B \partial_B
 \;.
\ee
The coefficient $\kappa$ satisfies the same constraint equation as before. The remaining coefficients $\xi_A$ are obtained from  (\ref{20XII11.1}), in notation adapted to the current setting:
\begin{eqnarray}
  - \frac{1}{2}(\partial_1 + \tau)\xi_A  + \tilde{\nabla}_B \sigma_A^{\phantom{A}B} - \frac{n-2}{n-1}\partial_A\tau -\partial_A \kappa
 = \overline T_{1A}
 \;.
 \label{20XII11.1xb}
\end{eqnarray}

The fact that this system of ODEs can be solved in a rather straightforward way (compare  (\ref{26I12.2})) given $\kappa$ and $g_{AB}$, should not prevent one to view this equation as a constraint on the initial data.

A useful observation here is that in~\cite[Appendix~C]{JKCPRD}, where it is shown that  (\ref{31III10.1}) and  (\ref{20XII11.1xb}) can be obtained from the usual \emph{vector constraint equation} on a spacelike hypersurface by a limiting process, when considering a family of spacelike hypersurfaces which become null in the limit.

The alert reader will note that the  constraint equations  (\ref{31III10.1}) and  (\ref{20XII11.1xb}) exhaust~all tangential components of $T_{\mu\nu}\ell^\nu$. We are not aware of a geometric interpretation of the equation    (\ref{30I12.1}), which involves the remaining, transverse, component of  $T_{\mu\nu}\ell^\nu$. Pursuing the analogy with the spacelike Cauchy problem, one could be  tempted to think of this equation as corresponding to the scalar spacelike constraint equation. However, this analogy is wrong since it is shown in~\cite[Appendix~C]{JKCPRD} that the scalar constraint equation and one of the vector constraint equations degenerate to the same single equation when a family of spacelike hypersurfaces degenerates to a characteristic one.

\subsection{Uniqueness of solutions}

Given a vacuum characteristic data set on a light-cone, or two vacuum characteristic data sets with a common boundary $S$ (where some further data might have to be prescribed, as made clear in previous sections), one can impose various supplementary conditions to construct an associated space-time metric. For example, one can  redefine $x^1$ so that $\kappa=0$ and impose wave-coordinate conditions, or wave-map coordinate conditions in the light-cone case, to obtain the required space-time metric. Or one can prescribe  the remaining metric functions as in Section~\ref{S28XII11.2}, with appropriate conditions at the tip of the light-cone or at the intersection surface. In~\cite[Section~7.1]{CCM2} a scheme is presented where $g_{12}|_{\mcN}$ is prescribed, together with wave-map conditions. It is obvious that there exist further schemes which are mixtures of the above and which might be more appropriate for some specific physical situations, or for matter fields with exotic coupling to gravity.

Rendall's analysis, or that in~\cite{CCM2}, makes it clear that every vacuum characteristic data set as  defined in Section~\ref{s26I11.1} leads to a unique, up to isometry, associated space-time, either near the tip of the light-cone, or near the intersection surface $S$. Here uniqueness is understood locally, though again it is clear that unique maximal globally hyperbolic developments should exist in the current context.
\section{Solving the constraint equation}
\label{solving_constraints}

There are several ways of solving (\ref{31III10.1}). The aim of this section is to present those methods, in vacuum. One should keep  in mind that some further specific hypotheses on the matter fields might have to be made in the schemes below for non-vacuum initial data:

\subsection{Solving for $\kappa$}
\label{s4V12.20}
 For any $\tilde g$ for which $\tau$ has no zeros,
      (\ref{29XII11.1})-(\ref{31III10.1}) can be solved algebraically for
     $\kappa$. This appears to be the most natural choice
     near the tip of a light-cone, where $\tau$ is nowhere vanishing.

\subsection{$\tau$ and $[\og_{AB}]$ as free data}

Another way of solving (\ref{31III10.1}) is to prescribe $[\og_{AB}]$ and the mean null extrinsic curvature $\tau$.
Here one can simply choose  $\tau$ to be nowhere vanishing such that  (\ref{31III10.1}) is solvable for $\kappa$.
Regularity conditions on $\tau$ in the light-cone-case are discussed in an appendix.

\subsection{$\kappa=0$}
\label{s4V12.12}
 Rendall's proposal is to reparameterize the characteristic curves so that $\kappa=0$, (\ref{31III10.1})  can then
     be rewritten as a linear equation for a conformal
     factor, say $\Omega$, such that $ g_{AB}= \Omega^2
     \gamma_{AB}$, where $\gamma_{AB}=[\og_{AB}]$ is freely
     prescribed; compare  (\ref{ODE_Omega}).

\subsection{$\kappa$ and $[\og_{AB}]$ as free data}
 \label{ss22I12.1}
In some situations it might be convenient not to assume that $\kappa=0$, but retain a version of the conformal approach of Rendall. This requires only a few minor modifications in Section~\ref{S28XII11.1}:
it suffices to replace (\ref{21XII11.30x}) by (\ref{21XII11.30}),  (\ref{20XII11.1x}) by (\ref{20XII11.1}),  (\ref{9.3}) by (\ref{9.3x}) and  (\ref{9.2xx}) by (\ref{constraint3})  with $\overline T_{\mu\nu}$, $\overline{\mathring W}{}^{\mu}$ and $\overline{\hat W}{}^{\mu}$  set to zero.
As a matter of course the corresponding equations on $N_2$ have to be adjusted analogously.

By an appropriate choice of $\kappa$, i.e.\ by choosing an adapted parameterization of the null rays, equation  (\ref{21XII11.30}),
which determines $\tau$, can sometimes be simplified; an example will be given in the next section.

\subsection{$\kappa=\tau/(n-1)$}
 \label{ss23II12.1}

An elegant approach is due to
     Hayward~\cite{HaywardNullSurfaceEquations} who, in
     space dimension $n=3$, proposes to use a parameterisation where $\kappa  = \tau/(n-1)$.
     Then, in vacuum, (\ref{31III10.1})
    becomes a linear
     equation for $\tau$, in terms of
     the trace-free part of $\chi$  which depends only upon
     the conformal class of  $\tilde g$,
\begin{equation}
  \partial_1\tau + |\sigma|^2 = 0
 \label{tau_kappa=tau-gauge}
 \;.
 \end{equation}
    The
     solution $\tau$ can then be used to determine   a
     conformal factor $\Omega^2$ relating $\og_{AB}$
     with a freely prescribed representative $\gamma_{AB}$ of the conformal
     class,
\begin{equation*}
 \partial_1\Omega - \frac{\Omega}{n-1}\left(\tau - \frac{1}{2}\gamma^{AB}\partial_1\gamma_{AB}\right) =0
 \;.
\end{equation*}

In the case where data are given on a light-cone, one has to face the question of boundary conditions for  (\ref{tau_kappa=tau-gauge}), of the (necessary and/or sufficient)  conditions on the data which will guarantee regularity at the vertex, and a possible relation between those.

To address those questions, we start by comparing the $\kappa=\tau/(n-1)$-gauge with the geometric $\mathring\kappa=0$ gauge. Those quantities computed in the latter gauge will be labelled by $\mathring{}$ in what follows.
Both  gauges are related by an angle-dependent rescaling of the coordinate $r$:
Using the transformation law of the Christoffel symbols we find
\begin{eqnarray}
 \tau/(n-1) \,= \,  \kappa \,= \, \Gamma^1_{11}\,= \,  \frac{\partial r}{\partial  \mathring r}\underbrace{\mathring \Gamma^1_{11}}_{=\mathring\kappa=0}   +   \frac{\partial r}{\partial  \mathring r}\frac{\partial^2  \mathring r}{\partial r^2}
\label{christoffel_trafo}
\end{eqnarray}
for $r =r(\mathring r,\mathring x^A)$. That yields
\begin{eqnarray}
\mathring r(r) = \int^r e^{\frac{1}{n-1}\int^{r_1} \tau(r_2)\mathrm{d}r_2 }   \mathrm{d}r_1 =  \int^r e^{-\frac{1}{n-1}\int^{r_1} \int^{r_ 2}|\sigma(r_3)|^2\mathrm{d}r_3\mathrm{d} r_2} \mathrm{d}r_1   ,
 \label{coord_trafo}
\end{eqnarray}
where we have suppressed any angle-dependence, and left unspecified any potential constants of integration.
This defines the desired local diffeomorphism.

As an example (and to obtain some intuition for this gauge scheme) consider the flat case where $|\sigma|^2\equiv 0$ and  for which we can compute everything explicitly.
There is no difficulty in determining the transformed data which satisfy $|\mathring \sigma|^2\equiv 0$.
The general solution of (\ref{tau_kappa=tau-gauge}) is
\begin{equation*}
 \tau(r,x^A) = \tau_0(x^A)
 \;.
\end{equation*}
Now we can explicitly compute (\ref{coord_trafo}),
\begin{equation}
  \mathring r(r) = A^{(1)} + A^{(2)} e^{\frac{\tau_0}{n-1}r}
 \label{trafo_Mink}
\end{equation}
for some integration functions $A^{(i)}$, with $A^{(2)}$ and $\tau_0$ nowhere vanishing since we seek a map $r\mapsto \mathring r$ which is a diffeomorphism on each generator. Then
\begin{equation}
 \mathring \tau(\mathring r) = \left(\frac{\partial \mathring r}{\partial  r}\right)^{-1} \tau(r(\mathring r)) = \frac{n-1}{\mathring r - A^{(1)}}
 \;.
 \label{eqn_tauring}
\end{equation}
We choose, as usual, the affine parameter $\mathring r$ in such a way that $\{\mathring r=0\}$ represents the vertex and such that $\mathring \tau = \frac{n-1}{\mathring r}$.
This leads to $A^{(1)}=0$.  Consequently, we either have to place the vertex at $r=-\infty$ and choose a positive $\tau_0$ or at $r=+\infty$ with a negative $\tau_0$ (w.l.o.g.\  we shall prefer the first alternative).
We conclude that in the $\kappa=\tau/(n-1)$-gauge we need to prescribe initial data for all $r\in\mathbb{R}$.

The regularity condition for $\mathring \tau$ translated into the $\kappa=\tau/(n-1)$-gauge does not lead to any boundary conditions for $\tau$, except for the requirement of constant sign.
It determines instead the position of the vertex, which in the new coordinates is located at infinity.

Let us come back to the general case, which we tackle from the other side, namely by starting in the $\mathring\kappa=0$-gauge. We use the identity
\( \frac{\partial^2\mathring r}{\partial r^2} = -\left(\frac{\partial \mathring r}{\partial r}\right)^3 \frac{\partial^2 r}{\partial \mathring  r^2} \) to rewrite (\ref{christoffel_trafo}):
\begin{eqnarray}
&&\frac{\partial \mathring r}{\partial r}\frac{\mathring \tau}{n-1} =  \frac{\tau}{n-1}= -  \left(\frac{\partial \mathring r}{\partial r}\right)^2 \frac{\partial^2 r}{\partial \mathring r^2}
 \nonumber
\\
&& \Longleftrightarrow \quad     \frac{\partial^2 r}{\partial \mathring r^2}  +  \frac{\mathring \tau}{n-1} \frac{\partial  r}{\partial \mathring r} =0
 \nonumber
\\
&&  \Longleftrightarrow \quad     r(\mathring r) = \int^{\mathring r} e^{-\frac{1}{n-1}\int^{\mathring r_1} \mathring \tau(\mathring r_2) \mathrm{d}\mathring r_2 }  \,\mathrm{d}\mathring r_1
 \;.
 \label{inverse_trafo}
\end{eqnarray}
This provides the inverse coordinate transformation.

Now, for a smooth metric, in adapted null coordinates which are constructed starting from normal coordinates the generators are affinely parameterized and we~have
\begin{equation}
 \mathring \tau = \frac{n-1}{\mathring r} + O(\mathring r)
 \;.
  \label{23II12.20}
\end{equation}
 But this behaviour remains unchanged under all reparameterisations of the generators which preserve the affine parameterisation
 as well as the position of the vertex. It follows that  (\ref{23II12.20}) holds for all smooth metrics in the gauge $\mathring \kappa =0$.

From  (\ref{inverse_trafo})-(\ref{23II12.20}) we obtain
\begin{equation*}
    r(\mathring r) = A^{(1)} + A^{(2)}\log\mathring r + O(\mathring r^2)\;, \quad A^{(2)}\ne 0 \enspace \forall \,x^A
 \;.
\end{equation*}
If we start in the $\mathring\kappa=0$-gauge, with the vertex at $\mathring r=0$, and transform into the $\kappa=\tau/(n-1)$-gauge then, similarly to Minkowski space-time, the vertex is shifted to,  w.l.o.g.,  $r=-\infty$.
Thus, space-time regularity forces the vertex  to be located at infinity in the $\kappa=\tau/(n-1)$-gauge.

\subsection{The shear as free data}
 \label{sssCfd}

Following Christodoulou~\cite{ChristodoulouBHFormation}, we let
the \emph{second fundamental form  $\chi$ of a null hypersurface ${\mcN}$ with null tangent $\ell$} be
defined as
\bel{secfunfo}
 \chi(X,Y)=g(\nabla_X \ell,Y)
 \;,
\ee
where $X,Y\in T{\mcN}$. Choosing $\ell$ to be $\partial_r$  we
then have, using \cite[Appendix~A]{CCM2},
\numparts
\bean
 \chi_{AB}
 &=&
 \overline{g(\nabla_A \partial_r,\partial_B)} = \og_{\mu B} \overline\Gamma^\mu _{Ar}
 = \overline g_{C B} \overline \Gamma^C _{Ar} + \overline g_{u B} \overline\Gamma^u
_{Ar}
\\
 & = & \frac 12 \partial_r \og_{AB}
  \;,
 \label{secfunfo2}
\\
 \chi_{rr}
 &=& \overline{
 g(\nabla_r \partial_r,\partial_r)}   = 0
  \;,
 \label{secfunfo4}
\\
 \chi_{Ar}
 &=& \overline{
 g(\nabla_A \partial_r,\partial_r)} = \og_{\mu r} \overline\Gamma^\mu _{Ar}
 = \overline g_{u r} \overline\Gamma^u _{rr} = 0
 \;.
\eeal{secfunfo3}
\endnumparts
Let $\sigma$ be the trace-free part of $\chi$ on the
level sets of $r$:
\begin{equation*}
\sigma_{AB}:= \chi_{AB} - \frac 1 {n-1} \og^{CD} \chi_{CD} \og_{AB}
 \;;
\end{equation*}
$\sigma$ is often called the \emph{shear} tensor of $\mcN$.
It has been proposed (cf., e.g.,~\cite{ChristodoulouBHFormation})  to
consider $\sigma$
as the free gravitational data at
${\mcN}$. There is an apparent problem with this proposal,
because to define a trace-free tensor one needs a
conformal metric; but if a conformal class $[\tg(r)]$
is given on
${\mcN}$, there does not seem to
be any need to supplement this class with $\sigma$.
This issue can be taken care of by working
in a frame formalism, as follows:

Let $\mcN$ be a $n$-dimensional manifold
threaded by a family of curves, which we call \emph{characteristic curves}, or \emph{generators}. We assume moreover that each curve is equipped with a connection: if, in local
coordinates, $\partial_r$ is tangent to the curves, then we let $\kappa$ denote the
corresponding connection coefficient, as in  (\ref{20I12.1}).

Suppose, for the moment, that $\mcN$ is a characteristic hypersurface embedded as the submanifold $\{u=0\}$ in a space-time $\mcM$. Choose some local coordinates so that $\partial_r$ is tangent to  the characteristic curves of $\mcN$.
Let $e_a$ denote a basis of $T\mcM$ along $\mcN$ such that
\bel{21I12.3}
 \nabla_r e_a = 0
 \;.
\ee

Let $S$ denote an $(n-1)$-dimensional submanifold of $\mcN$ (possibly, but not necessarily, its boundary) which intersects the generators transversally.
We will further require on $S$ that $e_0 $ is null, and that for $a=2,\ldots,n$ the family of vectors $e_a$ is orthonormal. These properties will then hold along all those generators that meet $S$.

Let $x^A$ be local coordinates on $S$, we propagate those along the generators of $\mcN$ to a neighbourhood $\mcU\subset \mcN$ of $S$ by requiring $\mcL_{\partial_r}x^A =0$.

We choose $e_1\sim\partial_r$ at $S$;  (\ref{20I12.1}) implies then that this will hold throughout $\mcU$:
\bel{21I12.4}
e_1 = e_1{}^r \partial_r \mbox{\ on $\mcU$.}
\ee

We choose the $e_a$'s, $a=2,\ldots n$, to be tangent to $S$; since $T\mcN$ coincides with $e_0^\perp$, the $e_a$'s, $a=2,\ldots n$ will remain tangent to $\mcN$:
\bel{21I12.5}
e_a = e_a{}^r \partial_r + e_a{}^B \partial_B \mbox{\ on $\mcU$, $a=2,\ldots,n$.}
\ee

On $S$ we choose the vector $e_0$ to be null, orthogonal to $S$, with
\bel{21I12.5x}
 g(e_0,e_1)=1
 \;.
\ee

Let $\{\theta^a\}_{a=0,1,\ldots,n}$, be a space-time coframe, defined on $\mcU\subset\mcN$, dual to the frame $\{e_a\}_{a=0,1,\ldots,n}$. From what has been said we have
\bel{21I12.1}
 g = \underbrace{ \theta^0 \otimes \theta^1 + \theta^1 \otimes \theta^ 0 +
  \theta^2 \otimes \theta^2 + \ldots+ \theta^n \otimes \theta^n}_{
  =: \eta_{ab}\theta^a \theta^b}
  \;.
\ee
By construction, the one-forms $\theta^a$ are covariantly constant along the generators of $\mcN$:
\bel{21I12.2}
 \nabla_r \theta^a = 0
 \;.
\ee
Here $\nabla$ is understood as the covariant-derivative operator acting on one-forms.

Again by construction $\theta^0$ annihilates $T\mcN$, thus $\theta^0 \sim \mathrm{d}u$ along $\mcN$:
\bel{21I12.7}
\theta^0 =\theta^0{}_u   \mathrm{d} u \mbox{\ on $\mcU$.}
\ee
We further note that
\bel{21I12.10}
\theta^a(\partial_r) =0    \mbox{\ on $\mcU$ for $a=2,\ldots,n$.}
\ee
To see this, recall that $\partial_r$ is orthogonal to $\partial_A$, hence
\beaa
 0 =g(\partial_r, \partial_A) = \eta_{ab}\theta^a(\partial_r) \theta^b (\partial_B)
 = \sum_{a=2}^N  \theta^a(\partial_r) \theta^a (\partial_B)
 \;.
\eeaa
The result follows now from the fact that the $(n-1)\times (n-1)$ matrix $(  \theta^a (\partial_B))_{a\ge 2}$ is non-degenerate.

We would like to calculate $\tilde g:= \og_{AB}\mathrm d x^A \mathrm d x^B$ using the covectors $\theta^a$. For this,  note that  (\ref{21I12.1}) and  (\ref{21I12.7})  imply
\bel{21I12.12}
 \og_{AB}
 = \sum_{a=2}^N  \theta^a(\partial_A) \theta^a (\partial_B)
 \;.
\ee
Thus, to determine $\tg$ it suffices to know the components $( \theta^a{}_B:= \theta^a (\partial_B))_{a\ge 2}$.
Now, using  (\ref{21I12.10}) together with \cite[Appendix~A]{CCM2} we have for $a\ge 2$
\bel{21I12.11}
 0 = \nabla_r \theta^a{}_B =
 \partial_r  \theta^a{}_B - \overline \Gamma^\mu{}_{rB} \theta^a{}_ \mu
 =
 \partial_r  \theta^a{}_B - \frac 12 \og^{AC}\partial_r g_{CB} \theta^a{}_ A
  \;.
\ee
There holds thus the following evolution equation for $(  \theta^a{}_B )_{a\ge 2}$:
\bel{21I12.14}
 \partial_r  \theta^a{}_B -  \og^{AC}\chi_{CB} \theta^a{}_ A =0
  \;,
\ee
where $\og^{AC}$ denotes the matrix inverse to $ \sum_{a=2}^N  \theta^a{}_A  \theta^a {}_B $.

Let, as before, $\ell=\partial_r$ and define
\bel{21I12.8}
 B_{\mu\nu}:= \nabla_\mu \ell_\nu
 \;.
\ee
Let $B_{ab}$ denote the frame-components of $B$:
\bel{21I12.9}
 B_{ab}= e_a{}^\mu e_b{}^\nu B_{\mu\nu}
 \quad
 \Longleftrightarrow
 \quad
 B_{\mu\nu} = \theta^a{}(\partial_\mu)\theta^b{}(\partial_\nu) B_{ab}
 \;.
\ee
It follows from the definition  (\ref{secfunfo}) that $\chi$ encodes the information on components of $B$ in directions tangent to $\mcN$:
\bel{26I12.4}
 \chi_{AB}=  B_{AB}
 \;,
 \quad
 \chi_{rr}= B_{rr} =0
 \;,
 \quad
 \chi_{Ar}= B_{Ar} =0
 \;.
\ee

The key distinction between $B$ and $\chi$ is, that $B$ has  components along directions transverse to $\mcN$, while $\chi$ hasn't. Also note that for $a,b\ge 1$ the frame components   $B_{ab}$ only involve the coordinate components of $B_{\mu\nu}$ tangential to $\mcN$, so the frame formula
\begin{equation*}
 \chi_{ab} = B_{ab} \;, \ a,b\ge 1,
\end{equation*}
is geometrically sensible.

The last two equations in  (\ref{26I12.4}) give
\bel{21I12.15}
 \chi_{AC}=\sum_{a,b=2}^n \theta^a{}_ A \theta^b {}_C B_{ab}
  \equiv  \sum_{a,b=2}^n \theta^a{}_ A \theta^b {}_C \chi_{ab}
 \;,
\ee
which allows us to rewrite  (\ref{21I12.14}) as
\bel{21I12.15x}
 \partial_r  \theta^a{}_B -  \sum_{b,c=2}^n \og^{AC} \theta^b{}_ B \theta^c {}_C\chi_{bc} \theta^a{}_ A =0
  \;.
\ee
Now, for $a,c\ge 2$,
\bel{30I12.2}
 \eta^{ac}= \og^{\mu\nu} \theta^a{}_\mu \theta^c{}_\nu
 = \og^{AC} \theta^a{}_A \theta^c{}_C
 \;,
\ee
which leads to
\bel{21I12.16}
 \partial_r  \theta^a{}_B - \sum_{b,c=2}^{n} \eta^{ac} \theta^b{}_ B  \chi_{bc}  =0
  \;.
\ee

This equation leads naturally to the following picture, assuming  for simplicity  vacuum Einstein equations. Consider, first, two null transversely intersecting hypersurfaces $\mcN_1$ and $\mcN_2$, with $\mcN_1\cap \mcN_2 =S$. For $a,b\ge 2$ let $\eta^{ab}$ be one when $a$ and $b$ coincide, and zero otherwise.
In addition to $\kappa$, the gravitational data on $\mcN=\mcN_1\cup\mcN_2$ can be encoded in a field of symmetric $\eta$-trace-free $(n-1)\times (n-1)$ matrices $ \sigma_{ab}$,   $a,b=2,\ldots,n$.
\begin{equation*}
   \sigma_{ab} =  \sigma_{ba}\;,
 \quad
 \eta^{ab}  \sigma_{ab} = 0
 \;.
\end{equation*}
Let $\tau$ be a solution of the equation
\begin{equation}
 (\partial_r -\kappa) \tau + \frac {\tau^2} {n-1} +|\sigma|_\eta^2=0\;,\enspace \mbox{where $ |\sigma|^2_\eta:= \eta^{ac} \eta^{bd} \sigma_{ab} \sigma_{cd}$}
 \;.
\label{sigma_eta_tau}
\end{equation}
There remains the freedom to prescribe $\tau\equiv g^{AB}\chi_{AB} =  \sum_{a,b=2}^n\eta^{ab} \chi_{ab} $ on $S$ (one such function for each surface $N_1$ and $N_2$).
Define
\bel{21I12.30}
 \chi_{ab} = \sigma_{ab} + \frac \tau {n-1} \eta_{ab}
 \;.
\ee
Solving  (\ref{21I12.16}) for $\theta^a{}_B$ along the generators of   $\mcN_1$ and $\mcN_2$, we can calculate $\og_{AB}$ on $\mcN$ from  (\ref{21I12.12}), as long as the determinant of the matrix $(\theta^a{}_B)_{a\ge 2}$ does not vanish (which will be the case in a neighbourhood of $S$).
Here one has the freedom of prescribing $\theta^a{}_B$ on $S$ for $a\geq 2$.
The characteristic constraint equation $R_{\mu\nu}\ell^\mu\ell^\nu=0$ holds by construction.
Indeed, the relation $\sigma_{ab} = e_a{}^Ae_b{}^B \sigma_{AB}$, $a,b\geq 2$, can be justified in an analogous manner as equation (\ref{21I12.15}).
That gives, using  (\ref{30I12.2}), with $a,b,c,d\geq 2$,
\begin{eqnarray*}
  |\sigma|_{\eta}^2  \equiv \eta^{ac}\eta^{bd}\sigma_{ab}\sigma_{cd} &=& \overline g^{AC} \theta^a{}_A\theta^c{}_C \overline g^{BD}\theta^b{}_B \theta^d{}_D e_a{}^Ee_b{}^F \sigma_{EF}e_c{}^G e_d{}^H\sigma_{GH}
\\
 &=&  \overline g^{AC} \overline g^{BD} \sigma_{AB}\sigma_{CD} \,\equiv\, |\sigma|^2
 \;,
\end{eqnarray*}
and the assertion follows immediately.

 We can now apply any of the methods described previously (e.g., Rendall's original method if $\kappa=0$) to
obtain a solution of the characteristic Cauchy problem to the future of $\mcN$.%
\footnote{Note that prescribing $\sigma_{ab}$ and $\kappa$ is equivalent to prescribe $\chi_{ab}$ as primary data. If we assume for simplicity that the $\eta$-trace of $\chi_{ab}$ is nowhere-vanishing, one can then determine  $\kappa$ from  (\ref{sigma_eta_tau}), and continue as described in the paragraph following (\ref{21I12.30}).
One could also consider this procedure in an adapted frame: From $\chi_{AB}$ one determines successively $\overline g_{AB}$, $\tau$ and $\kappa$. However, since the $\og_{AB}$-trace of $\chi$ is then not known a priori, it is not clear how to satisfy the constraint  (\ref{sigma_eta_tau}).  }

One should keep in mind the following: prescribing the data   $e_a{}^A|_S$, or equivalently $\theta^a{}_B|_S$, determines the metric $g_{AB}|_S$ on $S$. There is a supplementary freedom of making an $O(n-1)$-rotation of the frame;
\begin{equation*}
e_a{}^A|_S (x^A) \mapsto \omega^b{}_a (x^A) e_b{}^A|_S (x^A)
\;,
\end{equation*}
where the $\omega^b{}_a (x^A)$'s  are $O(n-1)$-matrices.
Any such rotation needs to be reflected in the $\sigma_{ab}$'s:
\begin{equation*}
\sigma_{ab}(r,x^A) \mapsto \omega^c{}_a  (x^A)\omega^d{}_b (x^A)\sigma_{cd} (r,x^A)
\;.
\end{equation*}
So, in this construction $\sigma_{ab}$ undergoes gauge-transformations which are constant along the generators, and are thus non-local in this sense.

In the case of a light-cone, the above construction can be implemented by first choosing an orthonormal coframe $\mathring \phi^a \equiv \mathring \phi^a{}_A dx^A$, $a\ge 2$, for the unit round metric $s_{AB}dx^A dx^B$ on $S^{n-1}$. The solutions $\theta^a:=r \phi^a \equiv r \phi^a{}_A dx^A$, $a\ge 2$, of  (\ref{21I12.16}) are then chosen as the unique solutions asymptotic to $r \mathring \phi_a $. It would be of interest to settle the question, ignored here, of sufficient and necessary conditions on $\sigma_{ab}$ so that the resulting initial data on the light-cone can be realized by restricting a smooth space-time metric to the light-cone.

\subsection{Friedrich's free data}
 \label{ss26I12.1xyz}

In~\cite{F1,FriedrichCMP86}, Friedrich proposes alternative initial data on
${\mcN}$, based on the identity
\footnote{This equation reduces to the usual Riccati equation (cf., e.g., \cite{galloway-nullsplitting}) satisfied by the null extrinsic curvature tensor when $\kappa=0$. We are grateful to Jos\'e-Maria Mart\'in-Garcia for providing the general version of that equation.}
\bea
  \partial_r^2 \og_{AB} -\kappa \partial_r  \og_{AB}
   -   \frac 12  \og^{CD} \partial_r  \og_{AC} \partial_r \og_{BD}
   =
        -  2  \overline{R}_{A r B r }
    \;;
\eeal{19X.4}
equivalently
\bea
  \partial_rB_{AB} -\kappa B_{AB}
   -      \og^{CD}B_{AC}B_{BD}
   =
        -    \overline{R}_{A r B r }
    \;.
\eeal{19X.4x}
Equation  (\ref{19X.4}) shows that  the
component $\overline{R}_{A r B r } $ of the Riemann tensor can be
calculated in terms of $\kappa$ and the field $ \og_{AB}$.

Alternatively, given the fields $\overline{C}_{A r B r } $, $\rho \equiv\overline T_{rr}$
 and $\kappa$, together with suitable boundary conditions, one can solve  (\ref{19X.4}) to determine $\og_{AB}$. So, in space-time dimension four, Friedrich~\cite{F1} proposes to
use frame components of $\overline{C}_{A r B r } $ as the free data on
${\mcN}$.
There is, however, a problem, in that  $\overline{C}_{A r B r } $ is traceless
\begin{equation*}
 0=   \overline g^{\mu\nu} \overline C_{\mu r \nu r}
 =\overline g^{AC} \overline C_{A r
 C r }
 \;.
\end{equation*}
So this condition has to be built-in into the formalism. But, as in the previous section,
the tracelessness condition does not seemingly make sense unless the inverse metric $g^{AB}$, or at least its conformal class, are known.

This issue can again be taken care of by a frame formalism,  whatever the dimension, as follows:
Let the orthonormal frame $ e_a$,
$a=0,\ldots,n $, and its dual coframe $\theta^a$ be as in the last section.  The property that the frame is parallel along the generators implies
\bel{20X8.8}
   \partial_r  e_a {}^B= -  \Gamma^B_{rC}e_a{}^C = -\og^{BA}B _{AC} e_a{}^C
  \ \mbox{for $a\ge 2$}
 \;.
\ee
We then have, for $a,b\ge 2$,
\bean
 \partial_r B_{ab} & = & \partial_r( e_a {}^\mu e_ b {} ^\nu B_{\mu\nu})
  \nonumber
\\
 & = &
 \partial_r( e_a {}^A) e_ b {} ^C B_{AC} +
  e_a {}^A  \partial_r(e_ b {} ^C) B_{AC} +
  e_a {}^A e_ b {} ^C \partial_r B_{AC}
  \nonumber
\\
 & = &
 -\og^{AD}B_{DE}e_a {}^E e_ b {} ^C B_{AC} -
  e_a {}^A  \og^{CD}B_{DE} e_ b {} ^E  B_{AC}
\nonumber
\\
 &&
    +
  e_a {}^A e_ b {} ^C \big(
   \kappa B_{AC}
  +      \og^{ED}B_{AE}B_{CD}
        -    \overline{R}_{A r C r }
        \big)  \nonumber
\\
 & = &
  -
  e_a {}^A e_ b {} ^E \og^{CD} B_{AC}B_{DE}
    +
   \kappa B_{ab}
        -    e_a {}^A e_ b {} ^B \overline{R}_{A rB r }
        \;.
\eeal{22I12.1}
Using
\begin{equation*}
 \og^{CD}= \eta^{cd} e_c{} ^C e_d {}^D = \sum_{c,d=2}^n\eta^{cd}e_c{} ^C e_d {}^D
\end{equation*}
we conclude that
\bean
 \partial_r B_{ab} & = &
  - \sum_{c,d=2}^n\eta^{cd}
    B_{ac}B_{db}
    +
   \kappa B_{ab}
        -    e_a {}^A e_ b {} ^B \overline{R}_{A rB r }
        \;.
\eeal{22I12.1x}
From the definition of the Weyl tensor in dimension $n+1$,
\bean C_{\mu\nu\sigma\rho} &:=& R_{\mu\nu\sigma\rho}-\frac 1{n-1}\left(g_{\mu\sigma}R_{\nu\rho}-
g_{\mu\rho}R_{\nu\sigma}-g_{\nu\sigma}R_{\mu\rho}+g_{\nu\rho}R_{\mu\sigma}\right)
\\
&& + \frac 1{ n (n-1)} R (g_{\mu\sigma}g_{\nu\rho}-g_{\mu\rho}g_{\nu\sigma})
 \;,
\eeal{26I12.10}
we find
\bean \overline C_{ArBr} &=& \overline R_{ArBr}
-\frac 1{n-1}\og_{AB}\overline R_{rr}
 \;.
\eeal{26I12.10x}

For $a,b\ge 2$ let
\begin{equation*}
 \psi_{ab} := e_a{}^A e_b {}^B \overline C_{A r B r}
\end{equation*}
represent the components of $ \overline C_{A r B r }$ in the current frame. Then
$\psi_{ab}$ is symmetric, with vanishing $\eta$-trace. We finally obtain the following equation
for $B_{ab}\equiv \chi_{ab}$, $a,b\ge2$:
\begin{eqnarray}
 (\partial_r -\kappa)\chi_{ab} & = &
  - \sum_{c,d=2}^n\eta^{cd}
    \chi_{ac}\chi_{db}
        -    \psi_{ab}  -\frac 1{n-1}\eta_{ab}\overline  T_{rr}
        \;.
\label{22I12.1x2}
\end{eqnarray}
This equation shows that $(\kappa, \psi_{ab})$
can be used as the free data for the gravitational field: Indeed, given $\kappa$, $\psi_{ab}$ and the component $\overline T_{rr}$ of the energy-momentum tensor, we can integrate   (\ref{22I12.1x2})
to obtain $\chi_{ab}$.
Note that by taking the $\eta$-trace of (\ref{22I12.1x2}) one recovers the constraint (\ref{sigma_eta_tau}) (here with $\overline T_{rr}$ possibly non-vanishing).

In the case of two transverse hypersurfaces the integration leaves the freedom of prescribing two tensors $\chi_{ab}$ on $S$, one corresponding to $N_1$ and another for $N_2$.
Then one proceeds as in the previous section to construct the remaining data on the initial surfaces, keeping in mind the further freedom to choose $\theta^a{}_B|_S$, $a\geq 2$, on $S$.

On a light-cone,  (\ref{22I12.1x2}) should be integrated with vanishing data at the vertex.

\ack{PTC acknowledges useful discussions with Yvonne Choquet-Bruhat, Helmut Friedrich, Jacek Jezierski, Jerzy Kijowski and Jos\'e Maria Mart\'in-Garc\'ia.}

\appendix

\section{The expansion $\mathbf{\tau}$ and the location of the vertex}
 \label{s13III12.1}

The Hayward gauge-condition of Section~\ref{ss23II12.1} has led us to the interesting conclusion, that in some gauge choices the vertex  of the light-cone will   be located at infinity.
This raises the question:  Under which conditions is the hypothesis, that the vertex is located at $r=0$, consistent with natural boundary conditions at the tip of the light-cone?
  
We start with the derivation of a necessary condition which needs to be imposed on the behaviour of the initial data in the $\kappa=\tau/(n-1)$-gauge near the vertex in order to be compatible with regularity.
It is known \cite{CCM2} that in a $\mathring \kappa=0$--gauge arising from normal coordinates the initial data have to be
 of the form
\begin{equation}
 \mathring\gamma_{AB} = \mathring r^2 s_{AB} + h_{AB}\;, \quad \mathrm{where} \quad h_{AB} = O_1(\mathring r^4)
 \;.
 \label{form_initial_data}
\end{equation}
Recall that we decorate with a circle the quantities corresponding to the gauge $\mathring\kappa=0$.
We want to work out how such data look like in the $\kappa=\tau/(n-1)$-gauge.
The coordinate transformation (\ref{inverse_trafo}), which defines a local diffeomorphism as long as $A^{(2)}$ does not change sign, reads
\begin{eqnarray}
     r(\mathring r) &=& A^{(1)} + A^{(2)}\log\mathring r +  f_h   \qquad A^{(2)}\ne 0 \enspace \forall \,x^A
 \nonumber
\\
 & =& \log\mathring r +  f_h
 \;,
 \label{specific_trafo}
\end{eqnarray}
where $f_h = O_1(\mathring r^2)$ is determined by $h_{AB}$ and where $A^{(i)}$ are integration functions. Here we have set $A^{(1)}=0$ and $A^{(2)}=1$, so that the $r$-coordinate in the $\kappa=\tau/(n-1)$-gauge is completely fixed, once $\mathring r$ has been chosen.

From (\ref{specific_trafo}) we  extract the behaviour of the inverse transformation,
\begin{equation*}
 \mathring r(r) = e^r + g_h\;, \quad g_h = O_1(e^{3r})
\end{equation*}
(where the symbol $O$ in connection with the $r$-coordinate refers to the limit $r \rightarrow -\infty$).
Now we can compute the overall form of the initial data,
\begin{eqnarray}
 \gamma_{AB}(r) = \mathring\gamma_{AB}(\mathring r(r)) = e^{2r}s_{AB} + k_{AB} \;, \quad \mathrm{where} \quad k_{AB} = O_1(e^{4r})
 \;.
 \label{initial_data_kappa=tau}
\end{eqnarray}
This implies
\begin{eqnarray}
|\sigma|^2 = -\frac{1}{4} \left( \partial_1\gamma^{AB}\partial_1\gamma_{AB} + \frac{(\gamma^{AB}\partial_1\gamma_{AB})^2}{n-1} \right)    \,=\,O(e^{2r})
 \label{sigma_for_kappa=tau}
 \;.
\end{eqnarray}

Note that, in contrast to the $\mathring \kappa=0$-gauge, $\tau$ remains bounded at the vertex for regular light-cone data of the form (\ref{initial_data_kappa=tau}), due to (\ref{tau_kappa=tau-gauge}) and (\ref{sigma_for_kappa=tau}).
 
Next, let us show that a bounded $\tau$ can only be  compatible with regularity when the vertex is located at infinity. For definiteness, we   consider initial data $\overline g_{\mu\nu}$ with nowhere vanishing $\tau\equiv \frac{1}{2} \overline g^{AB} \partial_1\overline g_{AB}$, within the scheme of  Section~\ref{S28XII11.2}.
Then $\kappa$ is computed algebraically via (\ref{20XII11.21a}) (and depends on the initial data).
Note that at this stage $\tau$ is a known function of $r$ which can be regarded as ``gauge part'' of the initial data.

By  calculations similar to those in  (\ref{christoffel_trafo})-(\ref{coord_trafo}) we can then obtain the coordinate $\mathring r$ relevant to the $\mathring\kappa=0$-gauge:
\begin{eqnarray}
 \mathring r(r) = \int^r e^{ \int^{r_1} \kappa(r_2) \mathrm{d}r_2 } \mathrm{d}r_1 \,
 \;,
 \label{coord_trafo2}
\end{eqnarray}
and transform all the fields to this gauge.

We have the identity
\begin{eqnarray}
 \label{13III12.5}
 \tau(r) = \frac{\partial\mathring r}{\partial r} \mathring\tau(\mathring r(r))
 \;, \quad
 \mathrm{where} \quad   \mathring \tau = \frac{n-1}{\mathring r} + O(\mathring r)
 \;,
\end{eqnarray}
since we assume regular light-cone data.
We consider the maximal range of $\mathring r$, near $\mathring r=0$, where $\mathring \tau$ is positive. It follows from  (\ref{coord_trafo2}) that   $\mathring r \mapsto r(\mathring r)$ is monotone there. Let us assume that this function is strictly increasing (the decreasing case is handled in a similar way), and let $(R_1,R_2)$ denote the corresponding range of $r$, with $-\infty\leq R_1 < R_2 \leq \infty$.
Then $\tau$ is positive on $(R_1,R_2)$ by  (\ref{13III12.5}).
If we choose $R_1<r_0<R_2$ such that $r^{-1}(r_0)=\mathring r_0 >0$,
from the last equation we find, for some ($x^B$-dependent) constant $A$,
\begin{eqnarray*}
  \int_{r_0}^{r(\mathring r)}\tau \mathrm{d}\tilde r =    \int_{\mathring r_0}^{\mathring r} \left( \frac{n-1}{\tilde{\mathring r}} + O(\tilde {\mathring r})\right)\mathrm{d}\tilde {\mathring r}   =  A + (n-1)\log\mathring r + O(\mathring r^2)
 \;.
\end{eqnarray*}
The right-hand side diverges to minus infinity at the vertex $\mathring r=0$, which is mapped to $R_1$. This gives
\begin{equation}
 \int_{R_1}^{r_0}\tau \mathrm{d}\tilde r = +\infty
 \;.
  \label{13III12.6}
\end{equation}
We conclude that any gauge in which $\tau$ is bounded will force the vertex to lie at infinity, for initial data which can be realized by a smooth space-time metric.

As another application of  (\ref{13III12.6}) we reconsider the $\kappa=\tau/(n-1)$-gauge. Let us denote by $\tau_0(x^A)$ the integration function  which arises in the associated constraint equation $\partial_1 \tau + |\sigma|^2=0$.
The exponential decay of $|\sigma|^2$ at a regular vertex, cf.\ equation  (\ref{sigma_for_kappa=tau}), is compatible with  (\ref{13III12.6})  only if $\tau_0$ is bounded away from zero.

Finally, consider initial data on $(0,\infty)$ with
\begin{eqnarray*}
 \overline g_{AB} &=& r^2 s_{AB} + O_1(r^4)
\;,
\end{eqnarray*}
as  in Section~\ref{ss10I12.1}. The function $\tau$ satisfies then
\begin{eqnarray}
 \label{initia_data_newgauge}
\tau &=& \frac{n-1}{r} + O(r)
\;,
\end{eqnarray}
which is, not unexpectedly, fully compatible with  (\ref{13III12.6}).

\section*{References}
\bibliographystyle{unsrt}
\bibliography{../references/hip_bib,%
../references/reffile,%
../references/newbiblio,%
../references/newbiblio2,%
../references/chrusciel,%
../references/bibl,%
../references/howard,%
../references/bartnik,%
../references/myGR,%
../references/newbib,%
../references/Energy,%
../references/dp-BAMS,%
../references/prop2,%
../references/besse2,%
../references/netbiblio,%
../references/PDE}

\def\polhk#1{\setbox0=\hbox{#1}{\ooalign{\hidewidth
  \lower1.5ex\hbox{`}\hidewidth\crcr\unhbox0}}} \def\cprime{$'$}
  \def\cprime{$'$}
\providecommand{\bysame}{\leavevmode\hbox to3em{\hrulefill}\thinspace}
\providecommand{\MR}{\relax\ifhmode\unskip\space\fi MR }
\providecommand{\MRhref}[2]{%
  \href{http://www.ams.org/mathscinet-getitem?mr=#1}{#2}
}
\providecommand{\href}[2]{#2}
\begin{thebibliography}{10}

\bibitem{BartnikIsenberg}
Bartnik R and Isenberg J 2004  The constraint equations {\it The Einstein  equations and the large scale behavior of gravitational fields} (Basel: Birkh\"auser) 1--38 

\bibitem{BBM}
Bondi H, van der Burg M G J and Metzner A W K 1962 Gravitational waves in   general relativity VII: Waves from axi--symmetric isolated systems {\it Proc.\   Roy.\ Soc.\ London A} {\bf 269}  21--52 

\bibitem{CaciottaNicoloI}
Caciotta G and Nicol{\`o} F 2005 Global characteristic problem for Einstein vacuum equations with small initial data I: The initial data constraints {\it Jour.\ Hyp.\ Differ.\ Equ.} {\bf 2}  no.~1 201--277  (arXiv:gr-qc/0409028)

\bibitem{CaciottaNicoloII}
Caciotta G and Nicol{\`o} F 2010 On a class of global characteristic problems for the Einstein  vacuum equations with small initial data {\it Jour.\ Math.\ Phys.} {\bf 51} 102503  (arXiv:gr-qc/0608038)

\bibitem{CagnacEinsteinCRAS2}
Cagnac F 1966 Probl\`eme de {C}auchy sur les hypersurfaces caract\'eristiques des \'equations d'Einstein du vide {\it C. R. Acad. Sci. Paris S\'er. A-B}  {\bf 262}  A1356--A1359 

\bibitem{CagnacEinsteinCRAS1}
Cagnac F 1966 Probl\`eme de {C}auchy sur les hypersurfaces caract\'eristiques des \'equations d'Einstein du vide {\it C. R. Acad. Sci.  Paris S\'er. A-B} {\bf 262} A1488--A1491 

\bibitem{YvonneLiouville}
Choquet-Bruhat Y 1971 Probl\`eme de {C}auchy pour le syst\`eme int\'egro diff\'erentiel d'{E}instein-{L}iouville {\it  Ann. Inst. Fourier (Grenoble)}  {\bf 21} 181--201 

\bibitem{YCB:GRbook}
Choquet-Bruhat Y 2009 {\it General relativity and the {E}instein equations} (Oxford: Oxford University Press) 

\bibitem{CCM4}
Choquet-Bruhat Y, Chru\'{s}ciel P T and Mart\'in-Garc\'ia J M 2010  An existence theorem for the Cauchy problem on a characteristic cone for the  Einstein equations with near-round analytic data  {\it Proceedings of  the Petrov 2010 Anniversary Symposium on General Relativity and Gravitation (Kazan)}
  (arXiv:1012.0777 [gr-qc])

\bibitem{CCM3}
Choquet-Bruhat Y, Chru\'{s}ciel P T and  Mart\'in-Garc\'ia J M 2011 An existence theorem for the Cauchy problem on a characteristic  cone for the Einstein equations {\it Cont.\ Math.} {\bf 554}  73--81, {\it Proceedings of ``Complex Analysis \& Dynamical Systems IV" (Nahariya) May   2009}  (arXiv:1006.5558 [gr-qc])

\bibitem{CCM2}
Choquet-Bruhat Y, Chru\'{s}ciel P T and Mart\'in-Garc\'ia J M 2011 The Cauchy problem on a characteristic cone for the Einstein  equations in arbitrary dimensions {\it Ann.\ H.\ Poincar\'e} {\bf 12}  419--482 (arXiv:1006.4467 [gr-qc])

\bibitem{ChoquetYork79}
Choquet-Bruhat Y and  York J W 1979 The {C}auchy problem {\it General  Relativity and Gravitation -- the {E}instein Centenary}  ed A Held  (New York: Plenum) 99--160

\bibitem{ChristodoulouBHFormation}
Christodoulou D 2008 {\it The Formation of Black Holes in General Relativity} (Z\"urich: EMS Publishing House) (arXiv:0805.3880 [gr-qc])

\bibitem{ChristodoulouMzH}
Christodoulou D and M{\"u}ller zum Hagen M 1981 Probl\`eme de valeur  initiale caract\'eristique pour des syst\`emes quasi lin\'eaires du second  ordre {\it C. R. Acad. Sci. Paris S\'er. I Math.} {\bf 293}  39--42 

\bibitem{ChJezierskiCIVP}
Chru\'{s}ciel P T and Jezierski J 2012 On free general relativistic initial data on the light cone {\it Jour.\ Geom.\ Phys.} {\bf 62}  578--593 (arXiv:1010.2098 [gr-qc])

\bibitem{DamourSchmidt}
Damour T and Schmidt B 1990 Reliability of perturbation theory in general  relativity {\it Jour.\ Math.\ Phys.} {\bf 31}  2441--2453 

\bibitem{Dautcourt}
Dautcourt G 1963 Zum charakteristischen {A}nfangswertproblem der {E}insteinschen {F}eldgleichungen  {\it Ann. Physik (7)} {\bf 12} 302--324 

\bibitem{DossaAHP}
Dossa M 2003 Probl\`emes de {C}auchy sur un cono\"\i de caract\'eristique  pour les \'equations d'{E}instein (conformes) du vide et pour les \'equations   de {Y}ang-{M}ills-{H}iggs {\it  Ann.\ H.~Poincar\'e} {\bf 4}  385--411 

\bibitem{ChBActa}
Four{\`e}s-Bruhat Y 1952 Th\'eor\`eme d'existence pour certains syst\`emes  d'\'equations aux d\'eriv\'ees partielles non lin\'eaires {\it  Acta Math.}  {\bf 88}  141--225

\bibitem{F1}
Friedrich H 1981  On the regular and the asymptotic characteristic initial value  problem for {E}instein's vacuum field equations {\it Proc.\ Roy.\ Soc.\ London  Ser.\ A} {\bf 375} 169--184 

\bibitem{F2}
Friedrich H 1981 The asymptotic characteristic initial value problem for {E}instein's vacuum field equations as an initial value problem for a  first-order quasilinear symmetric hyperbolic system {\it Proc.\ Roy.\ Soc.\  London Ser.\ A} {\bf 378} 401--421 

\bibitem{FriedrichCMP86}
Friedrich H 1986 On purely radiative space-times {\it Commun.\ Math.\ Phys.} {\bf 103} 35--65 

\bibitem{FriedrichAdS}
Friedrich H 1995 Einstein equations and conformal structure: existence of  anti--de {S}itter--type space-times {\it Jour.\ Geom. Phys.} {\bf 17}  125--184

\bibitem{FriedrichNagy}
Friedrich H and Nagy G 1998 The initial boundary value problem for {E}instein's vacuum field equation {\it  Commun.\ Math.\ Phys.} {\bf 201} 619--655

\bibitem{FriedrichRendall}
Friedrich H and Rendall A 2000 The {C}auchy problem for the {E}instein  equations {\it Einstein's field equations and their physical implications} {\it (Lecture Notes in Phys. vol 540)} ed B Schmidt (Berlin:  Springer) 127--223  

\bibitem{galloway-nullsplitting}
Galloway G J 2000 Maximum principles for null hypersurfaces and null  splitting theorems {\it Ann.\ H.\ Poincar\'e} {\bf 1} 543--567 

\bibitem{HaywardNullSurfaceEquations}
Hayward S A 1993 The general solution to the {E}instein equations on a null   surface {\it  Class.\ Quantum Grav.} \textbf{10}  773--778 

\bibitem{Jezierski:2003hh}
Jezierski J 2004 Geometry of null hypersurfaces  {\it Relativity  Today (Proceedings of the Seventh Hungarian Relativity Workshop, 2003)} ed I Racz (Budapest: Akademiai Kiado)  (arXiv:gr-qc/0405108)

\bibitem{JKCRMP}
Jezierski J, Kijowski J and Czuchry E 2000 Geometry of null-like surfaces  in general relativity and its application to dynamics of gravitating matter {\it  Rep.\ Math.\ Phys.} {\bf 46} 399--418 

\bibitem{JKCPRD}
Jezierski J, Kijowski J and Czuchry E 2002 Dynamics of a self-gravitating lightlike matter shell: a  gauge-invariant {L}agrangian and {H}amiltonian description {\it Phys.\ Rev.\ D  (3)} {\bf 65} 064036 

\bibitem{KRSWCMP}
Kreiss H-O, Reula O, Sarbach O and Winicour J 2009 Boundary conditions  for coupled quasilinear wave equations with application to isolated systems {\it Commun.\ Math.\ Phys.} {\bf 289} 1099--1129 

\bibitem{MzHSeifertCIVP}
M{\"u}ller zum Hagen H and Seifert H-J 1977 On characteristic  initial-value and mixed problems {\it Gen.\ Rel.\ Grav.} {\bf 8}  259--301 

\bibitem{Newman:Penrose}
Newman E T and Penrose R 1962 An approach to gravitational radiation by a  method of spin coefficients {\it Jour.\ Math.\ Phys.} {\bf 3}  566--578

\bibitem{Penrose}
Penrose R 1965 Zero rest mass fields including gravitation {\it Proc. Roy. Soc.  Lond.} {\bf A284} 159--203

\bibitem{PenroseCIVP}
Penrose R 1980 Null hypersurface initial data for classical fields of  arbitrary spin and for general relativity {\it  Gen.\ Rel.\ Grav.} {\bf 12} 225--264

\bibitem{RendallCIVP}
Rendall A D 1990 Reduction of the characteristic initial value problem to the {C}auchy problem and its applications to the {E}instein equations {\it Proc.\ Roy.\ Soc.\ London A} {\bf 427} 221--239 

\bibitem{SachsCIVP}
Sachs R K 1962 On the characteristic initial value problem in gravitational  theory {\it Jour.\ Math.\ Phys.} {\bf 3} 908--914

\bibitem{SarbachTiglio}
Sarbach O and Tiglio M 2012 Continuum and Discrete Initial-Boundary-Value Problems and Einstein's Field Equations  (arXiv:1203.6443 [gr-qc]) to appear in {\it Living Reviews in Relativity}

\bibitem{CalvinEV}
Tadmon C (2011) The Goursat Problem for the Einstein-Vlasov System: (I) The  Initial Data Constraints (arXiv:1109.6844 [gr-qc])

\end{thebibliography}

\end{document}